# Location-Based Reasoning about Complex Multi-Agent Behavior

**Adam Sadilek**                                    SADILEK@CS.ROCHESTER.EDU
**Henry Kautz**                                       KAUTZ@CS.ROCHESTER.EDU
*Department of Computer Science, University of Rochester*
*Rochester, NY 14627, USA*

## Abstract

Recent research has shown that surprisingly rich models of human activity can be learned from GPS (positional) data. However, most effort to date has concentrated on modeling single individuals or statistical properties of groups of people. Moreover, prior work focused solely on modeling actual successful executions (and not failed or attempted executions) of the activities of interest. We, in contrast, take on the task of understanding human interactions, attempted interactions, and intentions from noisy sensor data in a *fully relational multi-agent setting*. We use a real-world game of capture the flag to illustrate our approach in a well-defined domain that involves many distinct cooperative and competitive joint activities. We model the domain using Markov logic, a statistical-relational language, and learn a theory that jointly denoises the data and infers occurrences of high-level activities, such as a player capturing an enemy. Our unified model combines constraints imposed by the geometry of the game area, the motion model of the players, and by the rules and dynamics of the game in a probabilistically and logically sound fashion. We show that while it may be impossible to directly detect a multi-agent activity due to sensor noise or malfunction, the occurrence of the activity can still be inferred by considering both its impact on the future behaviors of the people involved as well as the events that could have preceded it. Further, we show that given a model of successfully performed multi-agent activities, along with a set of examples of failed attempts at the same activities, our system automatically learns an augmented model that is capable of recognizing success and failure, as well as goals of people's actions with high accuracy. We compare our approach with other alternatives and show that our unified model, which takes into account not only relationships among individual players, but also relationships among activities over the entire length of a game, although more computationally costly, is significantly more accurate. Finally, we demonstrate that explicitly modeling unsuccessful attempts boosts performance on other important recognition tasks.

## 1. Introduction

Our society is founded on the interplay of human relationships and interactions. Since every person is tightly embedded in our social structure, the vast majority of human behavior can be fully understood only in the context of the actions of others. Thus, not surprisingly, more and more evidence shows that when we want to model behavior of a person, the single best predictor is often the behavior of people in her social network. For instance, behavioral patterns of people taking taxis, rating movies, choosing a cell phone provider, or sharing music are best explained and predicted by the habits of related people, rather than by all the "single person" attributes such as age, race, or education (Bell, Koren, & Volinsky, 2007; Pentland, 2008).

In contrast to these observations, most research effort on activity recognition to date has concentrated on modeling single individuals (Bui, 2003; Liao, Fox, & Kautz, 2004, 2005), or statistical properties of aggregate groups of individuals (Abowd, Atkeson, Hong, Long, Kooper, & Pinkerton, 1997; Horvitz, Apacible, Sarin, & Liao, 2005), or combinations of both (Eagle & Pentland, 2006).

Notable exceptions to this "isolated individuals" approach includes the work of Kamar and Horvitz (2009) and Gupta, Srinivasan, Shi, and Davis (2009), where simple relationships among people are just starting to be explicitly considered and leveraged. For instance, Eagle and Pentland (2006) elegantly model the location of individuals from multi-modal sensory data, but their approach is oblivious to the explicit effects of one's friends, relatives, etc. on one's behavior. The isolated individuals approximations are often made for the sake of tractability and representational convenience. While considering individuals independently of each other is sufficient for some constrained tasks, in many interesting domains it discards a wealth of important information or results in an inefficient and unnatural data representation. On the other hand, decomposing a domain into a set of entities (representing for instance people, objects in their environment, or activities) that are linked by various relationships (e.g., is-a, has-a, is-involved-in) is a natural and clear way of representing data.

To address the shortcomings of nonrelational behavior modeling, we introduce the capture the flag domain (described below), and argue for a statistical-relational approach to learning models of multi-agent behavior from raw GPS data. The CTF dataset is on one hand quite complex and recorded by real-world sensors, but at the same time it is well-defined (as per the rules of the game), thereby allowing for an unambiguous evaluation of the results.

Being able to recognize people's activities and reason about their behavior is a necessary precondition for having intelligent and helpful machines that are aware of "what is going on" in the human-machine as well as human-human relationships. There are many exciting practical applications of activity recognition that have the potential to fundamentally change people's lives. For example, cognitive assistants that help people and teams be more productive, or provide support to (groups of) disabled individuals, or efficiently summarize a long complex event to a busy person without leaving out essential information. Other important applications include intelligent navigation, security (physical as well as digital), human-computer interaction, and crowdsourcing. All these applications and a myriad of others build on top of multi-agent activity recognition and therefore require it as a necessary stepping stone. Furthermore, as a consequence of the anthropocentrism of our technology, modeling human behavior plays—perhaps surprisingly—a significant role even in applications that do not directly involve people (e.g., unmanned space probes).

Furthermore, reasoning about human *intentions* is an essential element of activity recognition, since if we can recognize what a person (or a group of people) *wants* to do, we can proactively try to help them (or—in adversarial situations—hinder them). Intent is notoriously problematic to quantify (e.g., Baldwin & Baird, 2001), but we show that in the capture the flag domain, the notion is naturally captured in the process of learning the structure of failed activities. We all know perhaps too well that a successful action is often preceded—and unfortunately sometimes also followed—by multiple failed attempts. Therefore, reasoning about attempts typically entails high practical utility, but not just for their relatively high frequency. Consider, for example, a task of real-time analysis of a security video system. There, detecting that a person or a group of people (again, relations) *intend* to steal something is much more important and useful than recognizing that a theft has taken (or even is taking) place, because then it is certainly too late to entirely *prevent* the incident, and it may also be too late or harder to merely stop it. We believe that recognition of attempts in people's activities is a severely underrepresented topic in artificial intelligence that needs to be explored more since it opens a new realm of interesting possibilities.

Before we delve into the details of our approach in Sections 5 and 6, we briefly introduce the CTF dataset (Section 2), highlight the main contributions of our work (Section 3), and review background material (Section 4). We discuss related work, conclude, and outline future work in Sections 7, 8 and 9 respectively.

This paper incorporates and extends our previous work (Sadilek & Kautz, 2010a, 2010b).

## 2. Capture The Flag Domain

Imagine two teams—seven players each—playing capture the flag (CTF) on a university campus, where each player carries a consumer-grade global positioning system (GPS) that logs its location (plus noise) every second (see Figure 1). The primary goal is to enter the opponent's flag area. Players can be captured only while on enemy territory by being tagged by the enemy. Upon being captured, they must remain in place until freed (tagged by a teammate) or the game ends. The games involve many competitive and cooperative activities, but here we focus on (both successful and attempted) capturing and freeing. Visualization of the games is available from the first author's website.

We collected four games of CTF on a portion of the University of Rochester campus (about 23 acres) with Columbus V-900 GPS loggers (one per player) with 1 GB memory card each that were set to a sampling rate of 1 Hz. The durations of the games ranged approximately from 4 to 15 minutes.

Our work is not primarily motivated by the problem of annotating strategy games, although there are obvious applications of our results to sports and combat situations. We are, more generally, exploring relational learning and inference methods for recognizing *multi-agent activities* from location data. We accept the fact that the GPS data at our disposal is inherently unreliable and ambiguous for any one individual. We therefore focus on methods that *jointly and simultaneously* localize and recognize the high-level activities of groups of individuals.

Although the CTF domain doesn't capture all the intricacies of life, it contains many complex, interesting, and yet well-defined (multi-agent) activities. Moreover, it is based on extensive real-world GPS data (total of 40,000+ data points). Thus most of the problems that we are addressing here clearly have direct analogs in everyday-life situations that ubiquitous computing needs to address—imagine people going about their daily lives in a city instead of CTF players, and their own smart phones instead of GPS loggers.

One of the main challenges we have to overcome if we are to successfully model CTF is the severe noise present in the data. Accuracy of the GPS data varies from 1 to more than 10 meters. In open areas, readings are typically off by 3 meters, but the discrepancy is much higher in locations with tall buildings (which are present within the game area) or other obstructions. Compare the scale of the error with the granularity of the activities we concern ourselves with: both capturing and freeing involves players that are within reaching distance (less than 1 meter) apart. Therefore, the signal to noise ratio in this domain is daunting.

The error has a systematic component as well as a significant stochastic component. Errors between devices are poorly correlated, because subtle differences between players, such as the angle at which the device sits in the player's pocket, can dramatically affect accuracy. Moreover, since we consider multi-agent scenarios, the errors in individual players' readings can add up, thereby creating a large discrepancy between the reality and the recorded dataset. Because players can move freely through open areas, we cannot reduce the data error by assuming that the players move along road or walkways, as is done in much work on GPS-based activity recognition (e.g., Liao et al., 2004). Finally, traditional techniques for denoising GPS data, such as Kalman filtering, are of little help, due to the low data rate (1 sample per second) relative to the small amount of time required for a player to completely change her speed or direction.

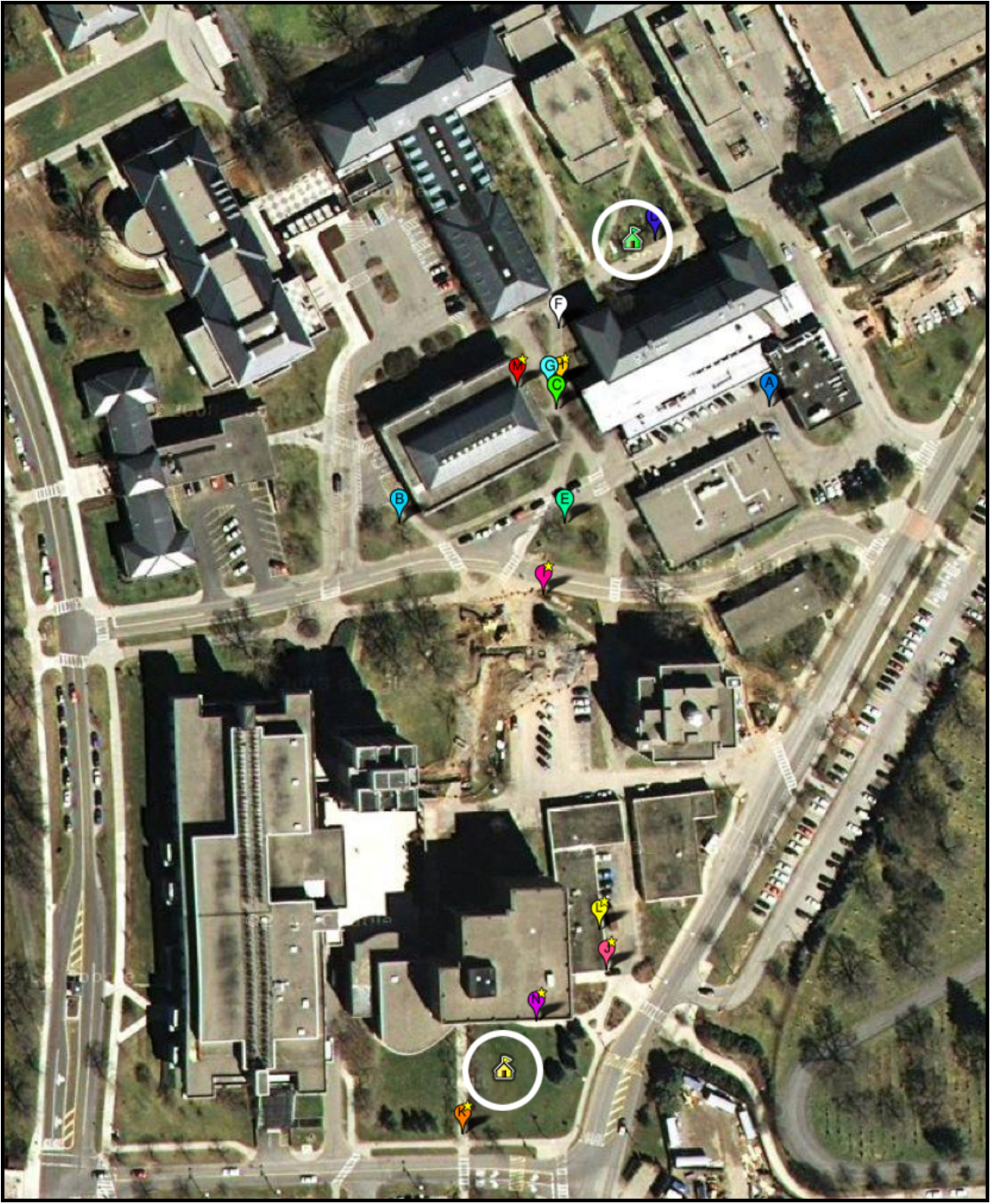

Figure 1: A snapshot of a game of capture the flag that shows most of the game area. Players are represented by pins with letters. In our version of CTF, the two "flags" are stationary and are shown as white circles near the top and the bottom of the figure. The horizontal road in the middle of the image is the territory boundary. The data is shown prior to any denoising or corrections for map errors. Videos of the games are available at `http://www.cs.rochester.edu/u/sadilek/`

If we are to reliably recognize events that happen in these games in the presence of such severe noise, we need to consider not only each player, but also the relationships among them and their actions over extended periods of time (possibly the whole length of the game). Consider a concrete task of inferring the individual and joint activities and intentions of the CTF players from their GPS traces. For example, suppose the GPS data shows player A running toward a stationary teammate B, then moving away. What occurred? Possibly player A has just "freed" player B, but GPS error has hidden the fact that player A actually *reached* B. Another possibility is that player A had the *intention* of freeing player B, but was scared off by an opponent at the last second. Yet another possibility is that no freeing occurred nor was even intended, because player B had not been previously captured.

Understanding a game thus consists of inferring a complex set of interactions among the various players as well as the players' intentions. The conclusions drawn about what occurs at one point in time affect and are affected by inferences about past and future events. In the example just given, recognizing that player B is moving in the future reinforces the conclusion that player A is freeing player B, while failing to recognize a past event of player B being captured decreases confidence in that conclusion. The game of CTF also illustrates that understanding a situation is as much or more about recognizing attempts and intentions as about recognizing successfully executed actions. For example, in course of a 15 minute game, only a handful of capture or freeing events occur. However, there are dozens of cases where one player unsuccessfully tries to capture an opponent or to free a teammate. A description of a game that was restricted to what actually occurred would be only a pale reflection of the original.

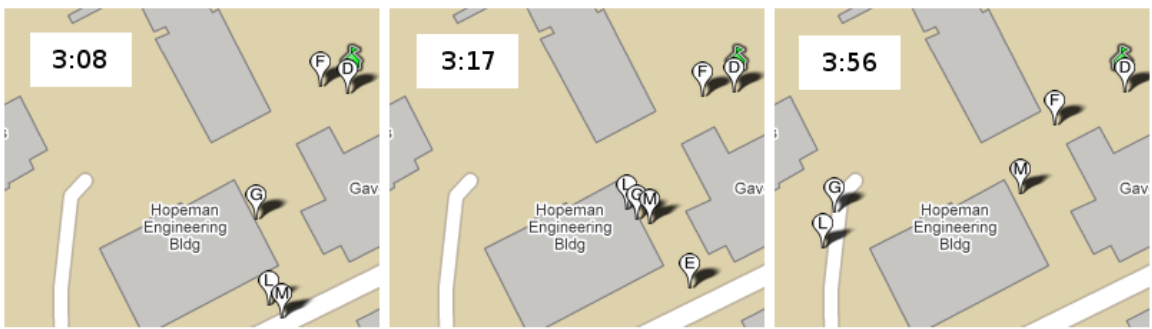

Figure 2: Three snapshots of a game situation where both successful and failed capturing occur. This example also illustrates the need for an approach that exploits both the relational and the far reaching temporal structure of our domain. (See text for explanation.)

As a concrete example, consider a real game situation illustrated in Figure 2. There we see three snapshots of a game projected over a map of the campus before any modification of the GPS data. The game time is shown on each snapshot. Players D, F, and G are allies and are currently on their home territory near their flag, whereas players L and M are their enemies. In the first snapshot, players L and M head for the opponent's flag but then—in the second frame—they are intercepted by G. At this point it is unclear what is happening because of the substantial error in the GPS data— the three players appear to be very close to each other, but in actuality they could have been 20 or more meters apart. However, once we see the third snapshot (note that tens of seconds have passed) we realize that player G actually captured only player M and didn't capture L since G is evidently still chasing L. The fact that player M remains stationary coupled with the fact that neither D nor F attempt to capture him suggests that M has indeed been captured. We show that it is possible to infer occurrences of capturing events even for complex situations like these whereas limited approaches largely fail. However, we need to be able to recognize not just individual events, we also need to discover new activities, identify their respective goals, and distinguish between events based on whether their outcomes are favorable or negative. For instance, in the second frame, player G tries to capture both L and M. Although he succeeded in the former case, he failed in the latter.

Many different kinds of cooperative and competitive multi-agent activities occur in the games. The lowest-level joint activities are based on location and movement, and include "approaching" and "being at the same location." Note, that noise in the GPS data often makes it difficult or impossible to directly detect these simple activities. At the next level come competitive multi-agent activities including capturing and attacking; cooperative activities include freeing; and there are activities, such as chasing and guarding, that may belong to either category or to both categories. There are also more abstract tactical activities, such as making a sacrifice, and overall strategies, such as playing defensively. In this paper, we concentrate on activities at the first two levels.

## 3. Our Contributions

The main contributions of this paper are as follows. We first present a novel method that simultaneously denoises positional data and learns a model of multi-agent activities that occur there. We subsequently evaluate the model on the CTF dataset and show that it achieves high accuracy in recognizing complex game events.

However, creating a model by manually writing down new rules or editing existing axioms is laborious and prone to introduction of errors or unnecessarily complex theories. Thus, we would like to automate this process by *learning* (or *inducing*) new axioms from training data. For people, it is much easier to provide or validate concrete examples than to directly modify a model. This leads us to our second contribution: We show how to automatically augment a preexisting model of (joint) activities so that it is capable of not only recognizing successful actions, but also identifies *failed attempts* at the same types of activities. This line of work also demonstrates that explicitly modeling attempted interactions in a unified way improves overall model performance.

As our third contribution, we demonstrate that the *difference* (discussed below) between the newly learned definitions of a failed activity and the original definition of the corresponding successful activity directly corresponds to the *goal* of the given activity. For instance, as per the rules of the capture the flag game, a captured player cannot move until freed. When our system induces the definition of failed capture, the new theory does not contain such a constraint on the movement of the almost-captured player, thereby allowing him to move freely.

## 4. Background

The cores of our models described below are implemented in Markov logic (ML), a statistical-relational language. In this section, we provide a brief overview of ML, which extends finite first-order logic (FOL) to a probabilistic setting. For a more detailed (and excellent) treatment of FOL,

ML, and inductive logic programming see the work of Shoenfield (1967), Domingos, Kok, Lowd, Poon, Richardson, and Singla (2008), and De Raedt and Kersting (2008), respectively.

In order to compare the Markov logic based models to alternative approaches, we consider a dynamic Bayesian network (DBN) model in the experiments below as one of our baselines. We therefore review relevant aspects of DBNs in this section as well.

### 4.1 Markov Logic

Given the inherent uncertainty involved in reasoning about real-world activities as observed through noisy sensor readings, we looked for a methodology that would provide an elegant combination of probabilistic reasoning with the expressive, relatively natural, and compact but unfortunately strictly true or false formulas of first-order logic. And that is exactly what Markov logic provides and thus allows us to elegantly model complex finite relational non-i.i.d. domains. A Markov logic network (MLN) consists of a set of constants $\mathcal{C}$ and of a set of pairs $\langle \mathcal{F}_i, w_i \rangle$ such that each FOL formula $\mathcal{F}_i$ has a weight $w_i \in \mathbb{R}$ associated with it. Optionally, each weight can be further scaled by a real-valued function of a subset of the variables that appear in the corresponding formula. Markov logic networks that contain such functions are called *hybrid* MLNs (Wang & Domingos, 2008).

A MLN can be viewed as a template for a Markov network (MN) as follows: the MN contains one node for each possible ground atom of MLN. The value of the node is 0 if the corresponding atom is *false* and 1 otherwise. Two nodes are connected by an edge if the corresponding atoms appear in the same formula. Thus, the MN has a distinct clique corresponding to each grounding of each formula. By $\mathcal{F}_i^{g_j}$ we denote the $j$-th grounding of formula $\mathcal{F}_i$. The MN has a feature value $f_{i,j}$ for each $\mathcal{F}_i^{g_j}$ such that

$$f_{i,j} = \begin{cases} 1 & \text{if } \mathcal{F}_i^{g_j} \text{ is } true \\ 0 & \text{otherwise} \end{cases}$$

Each weight $w_i$ intuitively represents the relative "importance" of satisfying (or violating, if the weight is negative) the corresponding formula $\mathcal{F}_i$. More formally, the weight scales the difference in log-probability between a world that satisfies $n$ groundings of the corresponding formula and one that results in $m$ true groundings of the formula, all else being equal (*cf.* Equation 1). Thus the problem of satisfiability is relaxed in MLNs. We no longer search for a satisfying truth assignment as in traditional FOL. Instead, we are looking for a truth assignment that maximizes the sum of the weights of all satisfied formulas.

The weights can be either specified by the knowledge base engineer or, as in our approach, learned from training data. That is, we provide the learning algorithm with labeled capture instances and pairs of raw and corresponding denoised trajectories along with labeled instances of game events and it finds an optimal set of weights that maximize the likelihood of the training data. Weight learning can be done in either generative or discriminative fashion. Generative training maximizes the joint probability of observed (evidence) as well as hidden (query) predicates, whereas discriminative learning directly maximizes the conditional likelihood of the hidden predicates given the observed predicates. Since prior work demonstrated that Markov network models learned discriminatively consistently outperform their generatively trained counterparts (Singla & Domingos, 2005), we focus on discriminative learning in our activity recognition domain.

Once the knowledge base with weights has been specified, we can ask questions about the state of hidden atoms given the state of the observed atoms. Let $X$ be a vector of random variables (one random variable for each possible ground atom in the MN) and let $\boldsymbol{\chi}$ be the set of all possible instantiations of $X$. Then, each $x \in \boldsymbol{\chi}$ represents a possible world. If $(\forall x \in \boldsymbol{\chi})[\Pr(X = x) > 0]$ holds, the probability distribution over these worlds is defined by

$$\Pr(X = x) = \frac{1}{Z} \exp\left( \sum_i w_i n_i \left( x_{\{i\}} \right) \right) \tag{1}$$

where $n_i(x_{\{i\}})$ is the number of true groundings of $i$-th formula with $w_i$ as its weight in a world $x$ and

$$Z = \sum_{x \in \boldsymbol{\chi}} \exp\left( \sum_i w_i n_i \left( x_{\{i\}} \right) \right) \tag{2}$$

Equation 1 can be viewed as assigning a "score" to each possible world and dividing each score by the sum of all scores over all possible worlds (the constant $Z$) in order to normalize.

Maximum *a posteriori* (MAP) inference in Markov logic given the state of the observed atoms reduces to finding a truth assignment for the hidden atoms such that the weighed sum of satisfied clauses is maximal. Even though this problem is in general #P-complete, we achieve reasonable run times by applying Cutting Plane MAP Inference (CPI) (Riedel, 2008). CPI can be thought of as a meta solver that incrementally grounds a Markov logic network, at each step creating a Markov network that is subsequently solved by any applicable method—such as MaxWalkSAT or via a reduction to an integer linear program. CPI refines the current solution by searching for additional groundings that could contribute to the objective function.

Up to this point, we have focused on *first-order* Markov logic. In first-order ML, each variable ranges over objects present the domain (e.g., apples, players, or cars). On the other hand, in finite *second-order* Markov logic, we variablize not only objects but also predicates (relations) themselves (Kok & Domingos, 2007). Our CTF model contains a predicate variable for each *type* of activity. For example, we have one variable *captureType* whose domain is {capturing, failedCapturing} and analogously for freeing events. When grounding the second-order ML, we ground all predicate variables as well as object variables. There has also been preliminary work on generalizing ML to be well-defined over infinite domains, which would indeed give it the full power of FOL (Singla & Domingos, 2007).

Implementations of Markov logic include Alchemy[1] and theBeast[2]. Our experiments used a modified version of theBeast.

## 4.2 Dynamic Bayesian Networks

A Bayesian network (BN) is a directed probabilistic graphical model (Jordan, 1998). Nodes in the graph represent random variables and edges represent conditional dependencies (*cf.* Figure 4). For a BN with $n$ nodes, the joint probability distribution is given by

$$\Pr(X_1, \ldots, X_n) = \prod_{i=1}^{n} \Pr\big(X_i | \mathrm{Pa}(X_i)\big), \tag{3}$$

---

1. `http://alchemy.cs.washington.edu/`

2. `http://code.google.com/p/theBeast/`

where $\mathrm{Pa}(X_i)$ denotes the parents of node $X_i$. In a typical setting, a subset of the random variables is *observed* (we know their actual values), while the others are *hidden* and their values need to be inferred.

A dynamic Bayesian network (DBN) is a BN that models sequential data. A DBN is composed of *slices*—in our case each slice represents a one second time interval. In order to specify a DBN, we either write down or learn intra- and inter-slice conditional probability distributions (CPDs). The intra-slice CPDs typically constitute the observation model while the inter-slice CPDs model transitions between hidden states. For an extensive treatment of DBNs, see the work of Murphy (2002).

There are a number of parameter learning and inference techniques for DBNs. To match the Markov logic-based framework, in the experiments with the DBN model presented below, we focus on a supervised learning scenario, where the hidden labels are known at training time and therefore a maximum likelihood estimate can be calculated directly.

We find a set of parameters (discrete probability distributions) $\theta$ that maximize the log-likelihood of the training data. This is achieved by optimizing the following objective function.

$$\theta^{\star} = \underset{\theta}{\mathrm{argmax}} \ \log\big(\mathrm{Pr}\big(x_{1:t}, y_{1:t}|\theta\big)\big), \tag{4}$$

where $x_{1:t}$ and $y_{1:t}$ represent the sequence of observed and hidden values, respectively, between times 1 and $t$, and $\theta^{\star}$ is the set of optimal model parameters. In our implementation, we represent probabilities and likelihoods with their log-counterparts to avoid arithmetic underflow.

At testing time, we are interested in the most likely explanation of the observed data. That is, we want to calculate the most likely assignment of states to all the hidden nodes (i.e., Viterbi decoding of the DBN) given by

$$y_{1:t}^{\star} = \underset{y_{1:t}}{\mathrm{argmax}} \ \log\big(\mathrm{Pr}(y_{1:t}|x_{1:t})\big), \tag{5}$$

where $\mathrm{Pr}(y_{1:t}|x_{1:t})$ is the conditional probability of a sequence of hidden states $y_{1:t}$ given a concrete sequence of observations $x_{1:t}$ between times 1 and $t$. We calculate the Viterbi decoding efficiently using dynamic programming (Jordan, 1998).

## 5. Methodology

In this section, we describe the three major components of our approach. In short, we first manually construct a model of captures and freeings in CTF and optimize its parameters in a supervised learning framework (Section 5.1). This constitutes our "seed" theory that is used for denoising raw location data and recognition of successful multi-agent activities. We then show, in Section 5.2, how to automatically extend the seed theory by inducing the structure and learning the importance of failed captures and freeings as well as the relationships to their successful counterparts. Finally, in Section 5.3, we use the augmented theory to recognize this richer set of multi-agent activities—both successful and failed attempts—and extract the goals of the activities.

Specifically, we investigate the following four research questions:

Q1. Can we reliably recognize complex multi-agent activities in the CTF dataset even in the presence of severe noise?

Q2. Can models of attempted activities be automatically learned by leveraging existing models of successfully performed actions?

Q3. Does modeling both success and failure allow us to infer the respective goals of the activities?

Q4. Does modeling failed attempts of activities improve the performance on recognizing the activities themselves?

We now elaborate on each of the three components of our system in turn, and subsequently discuss, in light of the experimental results and lessons learned, our answers to the above research questions.

## 5.1 Recognition of Successful Activities

In this section, we present our unified framework for intelligent relational denoising of the raw GPS data while simultaneously labeling instances of a player being captured by an enemy or freed by an ally. Both the denoising and the labeling are cast as a learning and inference problem in Markov logic. By denoising, we mean modifying the raw GPS trajectories of the players such that the final trajectories satisfy constraints imposed by the geometry of the game area, the motion model of the players, as well as by the rules and the dynamics of the game. In this paper, we refer to this trajectory modification as "snapping" since we tile the game area with 3 by 3 meter cells and snap each raw GPS reading to an appropriate cell. By creating cells only in unobstructed space, we ensure the final trajectory is consistent with the map of the area.

We begin by modeling the domain via a Markov logic theory, where we write the logical formulas that express the structure of the model by hand, and learn an optimal set of weights on the formulas from training data in a supervised discriminative fashion (details on the experimental set-up are in Section 6). In the following two subsections, we will show how to augment this seed Markov logic theory to recognize a richer set of events and extract the goals of players' multi-agent activities.

In order to perform data denoising and recognition of successful capturing and freeing, we model the game as weighted formulas in Markov logic. Some of the formulas are "hard," in the sense that we are only interested in solutions that satisfy all of them. Hard formulas capture basic physical constraints (e.g., a player is only at one location at a time) and inviolable rules of the game (e.g., a captured player must stand still until freed or the game ends).[3] The rest of the formulas are "soft," meaning there is a finite weight associated with each one. Some of the soft constraints correspond to a traditional low-level data filter, expressing preferences for smooth trajectories that are close to the raw GPS readings. Other soft constraints capture high-level constraints concerning when individual and multi-agent activities *are likely* to occur. For example, a soft constraint states that if a player encounters an enemy on the enemy's territory, the player is likely to be captured. The exact weights on the soft constraints are learned from labeled data, as described below.

We distinguish two types of atoms in our models: *observed* (e.g., GPS($P_1$, 4, 43.13°, −77.71°) and *hidden* (e.g., freeing($P_1$, $P_8$, 6)). The observed predicates in the CTF domain are: GPS, enemies, adjacent, onHomeTer, and onEnemyTer;[4] whereas capturing, freeing, isCaptured, isFree, samePlace, and snap are hidden. Additionally, the set of hidden predicates is expanded by the structure learning algorithm described below (see Table 1 for predicate semantics). In the training phase,

---

3. Cheating did not occur in our CTF games, but in principle could be accommodated by making the rules highly-weighted soft constraints rather than hard constraints.

4. While the noise in the GPS data introduces some ambiguity to the last two observed predicates, we can still reliably generate them since the road that marks the boundary between territories constitutes a neutral zone.

**Hard Rules:**

H1. Each raw GPS reading is snapped to exactly one cell.

H2.   (a) When player *a* frees player *b*, then both involved players must be snapped to a common cell at that time.

    (b) A player can only be freed by a free ally.

    (c) A player can be freed only when he or she is currently captured.

    (d) Immediately after a freeing event, the freed player transitions to a free state.

    (e) A player can only be freed while on enemy territory.

H3.   (a) When player *a* captures player *b*, then both involved players must be snapped to a common cell at that time.

    (b) A player can only be captured by a free enemy.

    (c) A player can be captured only if he or she is currently free.

    (d) Immediately after a capture event, the captured player transitions to a captured state.

    (e) A player can be captured only when standing on enemy territory.

H4. All players are free at the beginning of the game.

H5. At any given time, a player is either captured or free but not both.

H6. A player transitions from a captured state to a free state only via a freeing event.

H7. A player transitions from a free state to a captured state only via a capture event.

H8. If a player is captured then he or she must remain in the same location.

**Soft Rules:**

S1. Minimize the distance between the raw GPS reading and the snapped-to cell.

S2. Minimize projection variance, i.e., two consecutive "snappings" should be generally correlated.

S3. Maximize smoothness (both in terms of space and time) of the final player trajectories.

S4. If players *a* and *b* are enemies, *a* is on enemy territory and *b* is not, *b* is not captured already, and they are close to each other, then *a probably* captures *b*.

S5. If players *a* and *b* are allies, both are on enemy territory, *b* is currently captured and *a* is not, and they are close to each other, then *a probably* frees *b*.

S6. Capture events are generally rare, i.e., there are typically only a few captures within a game.

S7. Freeing events are also generally rare.

Figure 3: Descriptions of the hard and soft rules for capture the flag.

our learning algorithm has access to the known truth assignment to *all* atoms. In the testing phase, it can still access the state of the observed atoms, but it has to infer the assignment to the hidden atoms.

Figure 3 gives an English description of our hard and soft rules for the low-level movement and player interactions within capture the flag. Corresponding formulas in the language of ML are shown in Figures 5 and 6.

| Predicate | Type | Meaning |
|---|---|---|
| capturing$(a, b, t)$ | hidden | Player $a$ is capturing $b$ at time $t$. |
| enemies$(a, b)$ | observed | Players $a$ and $b$ are enemies. |
| adjacent$(c_1, c_2)$ | observed | Cells $c_1$ and $c_2$ are mutually adjacent, or $c_1 = c_2$. |
| failedCapturing$(a, b, t)$ | hidden | Player $a$ is unsuccessfully capturing $b$ at time $t$. |
| failedFreeing$(a, b, t)$ | hidden | Player $a$ is unsuccessfully freeing $b$ at time $t$. |
| freeing$(a, b, t)$ | hidden | Player $a$ is freeing $b$ at time $t$. |
| isCaptured$(a, t)$ | hidden | Player $a$ is in captured state at time $t$. |
| isFailedCaptured$(a, t)$ | hidden | At time $t$, player $a$ is in a state that follows an unsuccessful attempt at capturing $a$. $a$ in this state has the same capabilities as when free. |
| isFailedFree$(a, t)$ | hidden | At time $t$, player $a$ is in a state that follows an unsuccessful attempt at freeing $a$. $a$ in this state has the same capabilities as when captured. |
| isFree$(a, t)$ | hidden | Player $a$ is in free state at time $t$ (isFree$(a, t) \equiv \neg$ isCaptured$(a, t)$). |
| onEnemyTer$(a, t)$ | observed | Player $a$ in on enemy territory at time $t$. |
| onHomeTer$(a, t)$ | observed | Player $a$ in on home territory at time $t$. |
| samePlace$(a, b, t)$ | hidden | Players $a$ and $b$ are either snapped to a common cell or to two adjacent cells at time $t$. |
| snap$(a, c, t)$ | hidden | Player $a$ is snapped to cell $c$ at time $t$. |

Table 1: Summary of the logical predicates our models use. Predicate names containing the word "failed" are introduced by the Markov logic theory augmentation method described in Section 5.2.1.

We compare our unified approach with four alternative models. The first two models (**baseline** and **baseline with states**) are purely deterministic and they separate the denoising of the GPS data and the labeling of game events. We implemented both of them in Perl. They do not involve any training phase. The third alternative model is a **dynamic Bayesian network** shown in Figure 4. Finally, we have two models cast in Markov logic: the **two-step ML model** and the **unified ML model** itself. The unified model handles the denoising and labeling in a joint fashion, whereas the two-step approach first performs snapping given the geometric constraints and subsequently labels instances of capturing and freeing. The latter three models are evaluated using four-fold cross-validation where in order to test on a given game, we first train a model on the other three games.

All of our models can access the following observed data: raw GPS position of each player at any time and indication whether they are on enemy or home territory, location of each 3 by 3 meter cell, cell adjacency, and list of pairs of players that are enemies. We tested all five models on the same observed data. The following describes each model in more detail.

- **Baseline Model (B)**

  This model has two separate stages. First we snap each reading to the nearest cell and afterward we label the instances of player $a$ capturing player $b$. The labeling rule is simple:

we loop over the whole discretized (via snapping) data set and output capturing($a, b, t$) every time we encounter a pair of players $a$ and $b$ such that they were snapped (in the first step) to either the same cell or to two mutually adjacent cells at time $t$, they are enemies, and $a$ is on its home territory while $b$ is not. Freeing recognition is not considered in this simple model since we need to have a notion of persisting player states (captured or free) in order to model freeing in a meaningful way.

- **Baseline Model with States (B+S)**

  This second model builds on top of the previous one by introducing a notion that players have states. If player $a$ captures player $b$ at time $t$, $b$ enters a captured state (in logic, isCaptured($b, t + 1$)). Then $b$ remains in captured state until he moves (is snapped to a different cell at a later time) or the game ends. As per rules of CTF, a player who is in captured state cannot be captured again.

  Thus, this model works just like the previous one except whenever it is about to label a capturing event, it checks the states of the involved players and outputs capturing($a, b, t$) only if both $a$ and $b$ are *not* in captured state.

  Freeing recognition is implemented in an analogous way to capturing recognition. Namely, every time a captured player $b$ is about to transition to a free state, we check if $b$ has a free teammate $a$ nearby (again, within the adjacent cells). If that is the case, we output freeing($a, b, t$).

- **Dynamic Bayesian Network Model (DBN)**

  The dynamic Bayesian network model can be viewed as a probabilistic generalization of the above baseline model with states. The structure of the DBN model for one player is shown in Figure 4. In each time slice, we have one hidden node and four observed nodes, all of which represent binary random variables. We want to infer the most likely state $S$ for each player at any given time $t$ over the course of a game. The state is either free or captured and is hidden at testing time. There are four observed random variables per time step that model player's motion ($M$), presence or absence of at least one enemy ($EN$) and ally ($AN$) player nearby, and finally player's location on either home or enemy territory ($ET$). Each player is modeled by a separate DBN. Therefore, there are fourteen instantiated DBNs for each game, but within any one game, all the DBNs share the same set of parameters.

  Note that the DBN model does not perform any GPS trajectory denoising itself. To make a fair comparison with the Markov logic models, we use the denoising component of the Markov logic theory using only constraints H1 and S1–S3 (in Figure 3). This produces a denoised discretization of the data that is subsequently fed into the DBN model. The random variables within the DBN that capture the notion of player "movement" and players being "nearby" one another is defined on the occupancy grid of the game area, just like in the two deterministic baseline models. Namely, a player is said to be moving between time $t$ and $t + 1$ when he or she is snapped to two different nonadjacent cells at those times. Similarly, two players are nearby if they are snapped either to the same cell or to two adjacent cells.

- **Two-Step ML Model (2SML)**

  In the two-step approach, we have two separate theories in Markov logic. The first theory is used to perform a preliminary snapping of each of the player trajectories individually us-ing constraints H1 and S1–S3 (in Figure 3). This theory is identical to the one used in the discretization step in the DBN model above.

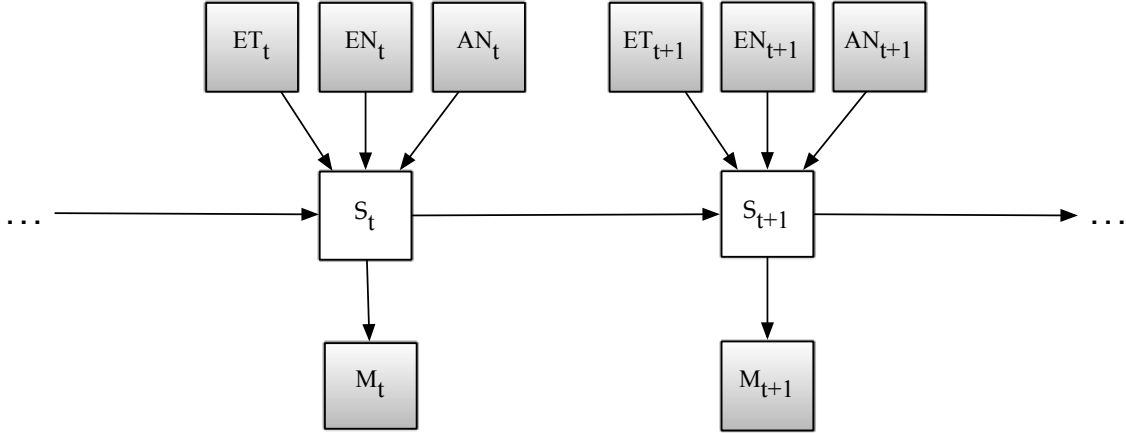

Figure 4: Two consecutive time slices of our dynamic Bayesian network for modeling the state of an individual player $P$ from observations. Shaded nodes represent observed random variables, unfilled denote hidden variables. All random variables are binary. ($ET_t = 1$ when $P$ is on enemy territory at time $t$, $EN_t = 1$ when there is an enemy nearby at time $t$, $AN_t = 1$ when there is an ally nearby at time $t$, and finally $M_t = 1$ if $P$ has moved between time $t-1$ and $t$. The value of hidden state $S_t$ is 1 if $P$ is captured at time $t$ and 0 when $P$ is free.)

The second theory then takes this preliminary denoising as a list of observed atoms in the form preliminarySnap($a, c, t$) (meaning player $a$ is snapped to cell $c$ at time $t$) and uses the remaining constraints to label instances of capturing and freeing, while considering cell adjacency in the same manner as the previous three models. The two-step model constitutes a decomposition of the unified model (see below) and overall contains virtually the same formulas, except 2SML operates with an observed preliminarySnap predicate, whereas the unified model contains a hidden snap predicate instead. Thus we omit elaborating on it further here.

- **Unified ML Model (UML)**

  In the unified approach, we express all the hard constraints H1–H8 and soft constraints S1–S7 (Figure 3) in Markov logic as a single theory that jointly denoises the data and labels game events. Selected interesting formulas are shown in Figure 6—their labels correspond to the listing in Figure 3. Note that formulas S1–S3 contain real-valued functions $d_1$, $d_2$, and $d_3$ respectively. $d_1$ returns the distance between agent $a$ and cell $c$ at time $t$. Similarly, $d_2$ returns the dissimilarity of the two consecutive "snapping vectors"[5] given agent $a$'s position at time $t$ and $t+1$ and the location of the centers of two cells $c_1$ and $c_2$. Finally, since people prefer to move in straight lines, function $d_3$ quantifies the lack of smoothness of any three consecutive segments of the trajectory. Since $w_p$, $w_s$, and $w_t$ are all assigned negative values during training, formulas S1–S3 effectively softly enforce the corresponding geometric constraints.

---

5. The initial point of each snapping (projection) vector is a raw GPS reading and the terminal point is the center of the cell we snap that reading to.

The presence of functions $d_1$ through $d_3$ renders formulas S1–S3 hybrid formulas. This means that at inference time, the instantiated logical part of each formula evaluates to either 1 (true) or 0 (false), which is in turn multiplied by the product of the corresponding function value and the formula weight.

We will see how we train, test, and evaluate these four models, and how they perform on the multi-agent activity recognition task in Section 6. Next, we turn to our supervised learning method for augmenting the unified ML model in order to recognize both successful and failed attempts at multi-agent activities.

---

**Hard formulas:**

$$\forall a, t \; \exists c : \mathrm{snap}(a, c, t) \tag{H1}$$

$$\forall a, c, c', t : (\mathrm{snap}(a, c, t) \wedge c \neq c') \Rightarrow \neg\mathrm{snap}(a, c', t)$$

$$\forall a_1, a_2, t : \mathrm{freeing}(a_1, a_2, t) \Rightarrow \big(\mathrm{samePlace}(a_1, a_2, t) \wedge \mathrm{isFree}(a_1, t) \wedge \tag{H2}$$
$$\neg\mathrm{enemies}(a_1, a_2) \wedge \mathrm{isCaptured}(a_2, t) \wedge \mathrm{isFree}(a_2, t + 1) \wedge$$
$$\mathrm{onEnemyTer}(a_1, t) \wedge \mathrm{onEnemyTer}(a_2, t)\big)$$

$$\forall a_1, a_2, t : \mathrm{capturing}(a_1, a_2, t) \Rightarrow \big(\mathrm{samePlace}(a_1, a_2, t) \wedge \mathrm{isFree}(a_1, t) \wedge \tag{H3}$$
$$\mathrm{enemies}(a_1, a_2) \wedge \mathrm{isFree}(a_2, t) \wedge \mathrm{isCaptured}(a_2, t + 1) \wedge$$
$$\mathrm{onHomeTer}(a_1, t) \wedge \mathrm{onEnemyTer}(a_2, t)\big)$$

$$\forall a_1, a_2, t : \mathrm{samePlace}(a_1, a_2, t) \Rightarrow \big(\exists c_1, c_2 : \mathrm{snap}(a_1, c_1, t) \wedge \mathrm{snap}(a_2, c_2, t) \wedge \mathrm{adjacent}(c_1, c_2)\big)$$

$$\forall a, t : (t = 0) \Rightarrow \mathrm{isFree}(a, t) \tag{H4}$$

$$\forall a, t : \mathrm{isCaptured}(a, t) \oplus \mathrm{isFree}(a, t) \tag{H5}$$

$$\forall a, t : (\mathrm{isFree}(a, t) \wedge \mathrm{isCaptured}(a, t + 1)) \Rightarrow (\exists_{=1} a_1 : \mathrm{capturing}(a_1, a, t)) \tag{H6}$$

$$\forall a, t : (\mathrm{isCaptured}(a, t) \wedge \mathrm{isFree}(a, t + 1)) \Rightarrow (\exists_{=1} a_1 : \mathrm{freeing}(a_1, a, t)) \tag{H7}$$

$$\forall a, t, c : (\mathrm{isCaptured}(a, t) \wedge \mathrm{isCaptured}(a, t + 1) \wedge \mathrm{snap}(a, c, t)) \Rightarrow \mathrm{snap}(a, c, t + 1) \tag{H8}$$

Figure 5: Our hard formulas in Markov logic. See corresponding rules in Figure 3 for an English description and Table 1 for explanation of the predicates. In our implementation, the actual rules are written in the syntax used by theBeast, a Markov logic toolkit. ($\exists_{=1}$ denotes unique existential quantification, $\oplus$ designates exclusive or.)

---

## 5.2 Learning Models of Failed Attempts

In the work described above, we manually designed the structure of a Markov logic network that models the capture the flag domain and allows us to jointly denoise the raw GPS data and recognize

**Soft formulas:**

$$\forall a, c, t : \big[\operatorname{snap}(a, c, t)\big] \cdot d_1(a, c, t) \cdot w_p \tag{S1}$$

$$\forall a, c_1, c_2, t : \big[\operatorname{snap}(a, c_1, t) \wedge \operatorname{snap}(a, c_2, t+1)\big] \cdot d_2(a, c_1, c_2, t) \cdot w_s \tag{S2}$$

$$\forall a, c_1, c_2, c_3, t : \big[\operatorname{snap}(a, c_1, t) \wedge \operatorname{snap}(a, c_2, t+1) \wedge \operatorname{snap}(a, c_3, t+2)\big] \cdot d_3(a, c_1, c_2, c_3, t) \cdot w_t \tag{S3}$$

$$\begin{aligned} \forall a_1, a_2, t : [&(\operatorname{enemies}(a_1, a_2) \wedge \operatorname{onHomeTer}(a_1, t) \wedge \\ &\operatorname{onEnemyTer}(a_2, t) \wedge \operatorname{isFree}(a_2, t) \wedge \\ &\operatorname{samePlace}(a_1, a_2, t)) \Rightarrow \operatorname{capturing}(a_1, a_2, t)] \cdot w_c \end{aligned} \tag{S4}$$

$$\begin{aligned} \forall a_1, a_2, t : [&(\neg \operatorname{enemies}(a_1, a_2) \wedge \operatorname{onEnemyTer}(a_1, t) \wedge \\ &\operatorname{onEnemyTer}(a_2, t) \wedge \operatorname{samePlace}(a_1, a_2, t) \wedge \operatorname{isFree}(a_1, t) \\ &\wedge \operatorname{isCaptured}(a_2, t)) \Rightarrow \operatorname{freeing}(a_1, a_2, t)] \cdot w_f \end{aligned} \tag{S5}$$

$$\forall a, c, t : \big[\operatorname{capturing}(a, c, t)\big] \cdot w_{cb} \tag{S6}$$

$$\forall a, c, t : \big[\operatorname{freeing}(a, c, t)\big] \cdot w_{fb} \tag{S7}$$

Figure 6: Soft formulas in Markov logic. See corresponding rules in Figure 3 for an English description. Each soft formula is written as a traditional quantified finite first-order logic formula (e.g., $\forall a, c, t : \big[\operatorname{snap}(a, c, t)\big]$), followed by an optional function (e.g., $d_1(a, c, t)$), followed by the weight of the formula (e.g., $w_p$). This syntax denotes that at inference time, the instantiated logical part of each formula evaluates to either 1 (true) or 0 (false), which is then effectively multiplied by the product of corresponding function value and formula weight.

instances of actual capturing and freeing. Now we show how to *automatically*—in a supervised learning setting—extend this theory to encompass and correctly label not only successful actions, but also failed attempts at those interactions. That is, given the raw GPS data that represent the CTF games, we want our new model to label instances where player $a$ captures (or frees) player $b$ as *successful captures* (*successful frees*) and instances where player $a$ almost captures (or frees) player $b$ as *failed captures* (*failed frees*). For example, by "failed capturing" we mean an instance of players' interactions where—up to a point—it appeared that $a$ is capturing $b$, but when we carefully consider the events that (potentially) preceded it as well as the impacts of the supposed capture on the future unfolding of the game, we conclude that it is a false alarm and no capture actually occurred. In other words, the conditions for a capture were right, but later on, there was a pivotal moment that foiled the capturing agent's attempt.

For both activities (capturing and freeing), our model jointly finds an optimal separation between success and failure. Note that since we cast our model in second-order Markov logic, we do not learn, e.g., an *isolated* rule that separates successful freeing from a failed attempt at freeing. Rather—since capturing and freeing events (both actual and failed) are related and thus labeling an activity as, say, "successful capturing" has far-reaching impact on our past, present, and future

labeling—we learn the separations in a joint and unified way. Namely, both the structure (logical form) and importance (weight) of each formula in our theory is considered with all its consequences and influence on other axioms in the theory. Our system thus finds an optimal balance between success and failure in capturing and freeing activities with respect to the training data.

### 5.2.1 THE THEORY AUGMENTATION ALGORITHM

In what follows, we will describe our Markov logic theory augmentation algorithm (Algorithm 1). For clarity, we will explain how it works in concrete context of the ML models of capture the flag we discussed in previous sections. However, the underlying assumption that successful actions are in many ways similar to their failed counterparts, and that minor—but crucial—deviations cause the failure to occur, often hold beyond capture the flag. Therefore, the same algorithm is applicable to other domains with different activities, as long as they are modeled in Markov logic.

---

**Algorithm 1** : Extend a ML theory to model successful as well as failed activities.

**Input:** $A$: set of activities

    $\mathcal{M}_S$: ML theory that models successful instances of activities in $A$

    $S$: set of examples of successful activities

    $F$: set of examples of failed activities

**Output:** $\mathcal{M}_{S+F}$: augmented ML model with learned weights that models both successful and attempted activities in $A$

    $\mathcal{I}$: intended goals of the activities

  1: $\mathcal{M}_S^2 \Leftarrow \text{liftToSecondOrderML}(\mathcal{M}_S, A)$

  2: $\mathcal{M}_S' \Leftarrow \text{instantiate}(\mathcal{M}_S^2, A)$

  3: $\mathcal{I} \Leftarrow \text{findIncompatibleFormulas}(F, \mathcal{M}_S')$

  4: $\mathcal{M}_{S+F} \Leftarrow \mathcal{M}_S' \setminus \mathcal{I}$

  5: $\mathcal{M}_{S+F} \Leftarrow \text{learnWeights}(S, F, \mathcal{M}_{S+F})$

  6: $\mathcal{M}_{S+F} \Leftarrow \text{removeZeroWeightedFormulas}(\mathcal{M}_{S+F})$

  7: **return** $\mathcal{M}_{S+F}, \mathcal{I}$

---

At a high-level, the augmentation algorithm belongs to the family of structure learning methods. Starting with a seed model of successful actions, it searches for new formulas that can be added to the seed theory in order to jointly model both successfully and unsuccessfully carried out actions. The declarative language bias—essentially rules for exploring the hypothesis space of candidate structures—is defined implicitly by the notion that for any given activity, the structure of unsuccessful attempts is similar to the successful attempts. Therefore, the augmentation algoritm goes through an "inflation" stage, where formulas in the seed theory are generalized, followed by a refinement stage, where superfluous and incompatible formulas in the inflated model are pruned away. The refinement step also optimizes the weights within the newly induced theory. We will now discuss this process in more detail.

The input of our theory augmentation algorithm consists of an initial first-order ML theory $\mathcal{M}_S$ that models successful capturing and freeing (such as the unified ML model defined in Section 5.1 that contains formulas shown in Figures 5 and 6), a set of activities of interest $A$, and a set of examples of successful ($S$) as well as failed ($F$) captures and frees. $\mathcal{M}_S$ does not need to have weights for its soft formulas specified. In case they are missing, we will learn them from scratch in the final steps of the augmentation algorithm. If the weights are specified, the final weight learning step for $\mathcal{M}_{S+F}$ can leverage them to estimate the initial weight values. $A$ can be specified as a set of predicate names, e.g., {capturing, freeing}. Each example in sets $S$ and $F$ describes a game segment and constitutes a truth assignment to the appropriate literals instantiated from $\mathcal{M}_S$. Table 2 shows two toy examples of sets $S$ and $F$ for three time steps. Since the goal is to learn a model of failed (and successful) attempts in a supervised way, the example game segment in $F$ contain activities labeled with predicates failedCapturing() and failedFreeing().

If $\mathcal{M}_S$ contains hybrid formulas (such our formulas S1–S3 in Figure 6), the appropriate function definitions are provided as part of $S$ and $F$ as well. Each definition consists of implicit mapping from input arguments to function values. For instance, function $d_1$ in formula S1 quantifies the L2 distance between the agent $a$ and cell $c$ at time $t$ in the projected Mercator space: $d_1(a, c, t) = \sqrt{(a.gpsX_t - c.gpsX)^2 + (a.gpsY_t - c.gpsY)^2}$.

Our system goes through the following process in order to induce a new theory $\mathcal{M}_{S+F}$ that augments $\mathcal{M}_S$ with a definition of failed attempts for each activity already defined in $\mathcal{M}_S$.

First we lift $\mathcal{M}_S$ to second-order Markov logic by variabilizing all predicates that correspond to the activities of interest (step 1 of Algorithm 1). This yields a lifted theory $\mathcal{M}_S^2$. More concretely, in order to apply this technique in our domain, we introduce new predicate variables *captureType* (whose domain is {capturing, failedCapturing}), *freeType* (over {freeing, failedFreeing}), and *stateType* (over {isCaptured, isFailedCaptured, isFree, isFailedFree}). For instance, variabilizing a first-order ML formula freeing$(a, b, t) \Rightarrow \neg$enemies$(a, b)$ yields a second-order ML formula *freeType*$(a, b, t) \Rightarrow \neg$enemies$(a, b)$ (note that *freeType* is now a variable). Instantiating back to first-order yields two formulas: freeing$(a, b, t) \Rightarrow \neg$enemies$(a, b)$ and failedFreeing$(a, b, t) \Rightarrow \neg$enemies$(a, b)$.

As far as agents' behavior is concerned, in the CTF domain, isCaptured is equivalent to isFailedFree, and isFree is equivalent to isFailedCaptured. As we will soon see, the theory augmentation process learns these equivalence classes and other relationships between states from training examples by expanding and subsequently refining formula H5 in Figure 5. While we could work with only the isCaptured predicate and its negation to represent agents' states, we feel that having explicit failure states makes our discussion clearer. Furthermore, future work will need to address hierarchies of activities, including their failures. In that context, a representation of explicit failure states may not only be convenient, but may be necessary.

Next, we instantiate all predicate variables in $\mathcal{M}_S^2$ to produce a new first-order ML theory $\mathcal{M}_S'$ that contains the original theory $\mathcal{M}_S$ in its entirety plus new formulas that correspond to failed captures and frees (step 2). Since events that are, e.g., near-captures appear similar to actual successful captures, our hypothesis is that we do not need to drastically modify the original "successful" formulas in order to model the failed activities as well. In practice, the above process of lifting and instantiating indeed results in a good seed theory. While we could emulate the lifting and grounding steps with a scheme of copying formulas and renaming predicates in the duplicates appropriately, we cast our approach in principled second-order Markov logic, which ties our work more closely to previous research and results in a more extensible framework. Specifically, second-order Markov logic has been successfully used in deep transfer learning (Davis & Domingos, 2009) and predicate invention (Kok & Domingos, 2007). Therefore, an interesting direction of future work is to combine our theory augmentation and refinement with transfer and inductive learning—operating in second-order ML—to jointly induce models of failed attempts of different activities in different domains, while starting with a single model of only successful activities in the source domain.

| Set *S*: Successful Capture | Set *F*: Failed Capture |
|---|---|
| enemies($P_1, P_2$) | enemies($P_4, P_5$) |
| enemies($P_2, P_1$) | enemies($P_5, P_4$) |
|  | onEnemyTer($P_5, 1$) |
| onEnemyTer($P_2, 2$) | onEnemyTer($P_5, 2$) |
| onEnemyTer($P_2, 3$) | onEnemyTer($P_5, 3$) |
| capturing($P_1, P_2, 2$) | failedCapturing($P_4, P_5, 2$) |
| isFree($P_1, 1$) | isFree($P_4, 1$) |
|  | isFailedCaptured($P_4, 1$) |
| isFree($P_1, 2$) | isFree($P_4, 2$) |
|  | isFailedCaptured($P_4, 2$) |
| isFree($P_1, 3$) | isFree($P_4, 3$) |
|  | isFailedCaptured($P_4, 3$) |
| isFree($P_2, 1$) | isFree($P_5, 1$) |
|  | isFailedCaptured($P_5, 1$) |
| isFree($P_2, 2$) | isFree($P_5, 2$) |
|  | isFailedCaptured($P_5, 2$) |
| isCaptured($P_2, 3$) | isFree($P_5, 3$) |
|  | isFailedCaptured($P_5, 3$) |
| snap($P_1, C5, 1$) | snap($P_4, C17, 1$) |
| snap($P_1, C10, 2$) | snap($P_4, C34, 2$) |
| snap($P_1, C10, 3$) | snap($P_4, C0, 3$) |
| snap($P_2, C9, 1$) | snap($P_5, C6, 1$) |
| snap($P_2, C10, 2$) | snap($P_5, C34, 2$) |
| snap($P_2, C10, 3$) | snap($P_5, C7, 3$) |
| samePlace($P_1, P_2, 2$) | samePlace($P_4, P_5, 2$) |
| samePlace($P_2, P_1, 2$) | samePlace($P_5, P_4, 2$) |
| samePlace($P_1, P_2, 3$) |  |
| samePlace($P_2, P_1, 3$) |  |

Table 2: Two examples of a logical representation of successful ($S$) as well as failed ($F$) capture events that are input to Algorithm 1. The closed-world assumption is applied, therefore all atoms not listed are assumed to be false. For clarity, we omit listing the adjacent() predicate.

Typical structure learning and inductive logic programming techniques start with an initial (perhaps empty) theory and iteratively grow and refine it in order to find a form that fits the training data well. In order to avoid searching the generally huge space of hypotheses, a declarative bias is either specified by hand or mined from the data. The declarative bias then restricts the set of possible refinements of the formulas that the search algorithm can apply. Common restrictions include limiting formula length, and adding a new predicate to a formula only when it shares at least one variable with some predicate already present in the formula. On the other hand, in our approach, we first generate our seed theory by instantiating all the activity-related predicate variables. To put it into context of structure learning, we expand the input model in order to generate a large seed theory, and then apply bottom-up (data-driven) learning to prune the seed theory, whereby the training data guides our search for formulas to remove as well as for an optimal set of weights on the remaining formulas. We conjecture that any failed attempt at an activity always violates at least one constraint that holds for successful executions of the activity. The experiments below support this conjecture.

The pruning is done in steps 3 and 4 of Algorithm 1. The function findIncompatibleFormulas($F$, $\mathcal{M}'_S$) returns a set of hard formulas in $\mathcal{M}'_S$ that are *incompatible* with the set of examples of failed interactions $F$. We say that a formula $c$ is compatible with respect to a set of examples $F$ if $F$ logically entails $c$ ($F \models c$). Conversely, if $F$ does not entail $c$, we say that $c$ is incompatible w.r.t. $F$. We explain how to find incompatible formulas in the next section.

In step 4 of Algorithm 1, we simply remove all incompatible formulas ($\mathcal{I}$) from the theory. At this point, we have our $\mathcal{M}_{S+F}$ model, where hard formulas are guaranteed logically consistent with the examples of failed activities (because we removed the incompatible hard formulas), as well as with the successful activities (because they were logically consistent to start with). However, the soft formulas in $\mathcal{M}_{S+F}$ are missing properly updated weights (in Markov logic, the weight of each hard formula is simply set to $+\infty$). Therefore, we run Markov logic weight learning using theBeast package (step 5).

Recall that theBeast implements the cutting plane meta solving scheme for inference in Markov logic, where the ground ML network is reduced to an integer linear program that is subsequently solved by the LpSolve ILP solver. We chose this approach as opposed to, e.g., MaxWalkSAT that may find a solution that is merely locally optimal, since the resulting run times are still relatively short (under an hour even for training and testing even the most complex model). Weights are learned discriminatively, where we directly model the posterior conditional probability of the hidden predicates given the observed predicates. We set theBeast to optimize the weights of the soft formulas via supervised on-line learning using margin infused relaxed algorithm (MIRA) for weight updates while the loss function is computed from the number of false positives and false negatives over the hidden atoms. Note that if any of the soft formulas are truly irrelevant with respect to the training examples, they are not picked out by the findIncompatibleFormulas() function, but their weights are set to zero (or very close to zero) in the weight learning step (line 5 in Algorithm 1). These zero-weighted formulas are subsequently removed in the following step. Note that the weight learning process does not need to experience a "cold" start, as an initial setting of weights can be inherited from the input theory $\mathcal{M}_S$.

Finally, we return the learned theory $\mathcal{M}_{S+F}$, whose formulas are optimally weighted with respect to *all* training examples. In the Experiments and Results section below, we will use $\mathcal{M}_{S+F}$ to recognize both successful and failed activities. Algorithm 1 also returns the incompatible hard formulas $\mathcal{I}$. We will see how $\mathcal{I}$ is used to extract the intended goal of the activities in the Section 5.3, but first, let us discuss step 3 of Algorithm 1 in more detail.

### 5.2.2 CONSISTENCY CHECK: FINDING INCOMPATIBLE FORMULAS

Now we turn to our method for finding incompatible formulas (summarized in Algorithm 2). Since our method leverages satisfiability testing to determine consistency between candidate theories and possible worlds (examples),[6] Algorithm 2 can be viewed as an instance of learning from interpretations—a learning setting in the inductive logic programming literature (De Raedt, 2008).

---

6. This is often referred to as the *covers* relation in inductive logic programming.

---

**Algorithm 2 (findIncompatibleFormulas).** Find formulas in a ML theory that are logically inconsistent with examples of execution of failed activities.

---

**Input:** $F$: a set of examples of failed activities

　　　　$\mathcal{T}$: unrefined ML theory of successful and failed activities

**Output:** smallest set of formulas that appear in $\mathcal{T}$ and are unsatisfiable in the worlds in $F$

1: $O \Leftarrow$ extractObjects$(F)$
2: $\mathcal{T}_{\text{hard}} \Leftarrow \mathcal{T} \setminus \mathcal{T}_{\text{soft}}$
3: **integer** $n \Leftarrow 0$
4: **boolean** result $\Leftarrow$ **false**
5: **while** result **== false do**
6: 　$\mathcal{T}^c \Leftarrow \mathcal{T}_{\text{hard}}$
7: 　remove a new $n$-tuple of formulas from $\mathcal{T}^c$
8: 　**if** for the current $n$, all $n$-tuples have been tested **then**
9: 　　$n \Leftarrow n + 1$
10: 　**end if**
11: 　result $\Leftarrow$ testSAT$(F, \mathcal{T}^c, O)$
12: **end while**
13: **return** $\mathcal{T}_{\text{hard}} \setminus \mathcal{T}^c$

---

As input, we take a set of examples of failed activities $F$ and a seed theory $\mathcal{T}$ (e.g., produced in step 2 of Algorithm 1). The output is the smallest set of hard formulas that appear in $\mathcal{T}$ and are logically inconsistent with $F$. The algorithm first extracts the set of all objects $O$ that appear in $F$ (step 1 in Algorithm 2), while keeping track of the type of each object. For example, suppose there are only two example worlds in $F$ shown in Table 3. Then *extractObjects(F)* returns $\{P_1, P_2, P_7, P_8, C_3, C_5, 1, 2\}$.

| **Example 1** | **Example 2** |
|---|---|
| snap$(P_1, C_5, 1)$ | snap$(P_7, C_3, 2)$ |
| snap$(P_2, C_5, 1)$ | snap$(P_8, C_3, 2)$ |
| failedCapturing$(P_1, P_2, 1)$ | failedFreeing$(P_2, P_5, 2)$ |

Table 3: Two simple examples of a logical representation a failed capture event.

In step 2, we limit ourselves to only hard formulas when testing compatibility. We do so since we can *prove* incompatibility only for hard formulas. Soft constraints can be violated many times in the data and yet we may not want to eliminate them. Instead, we want to merely adjust their weights, which is exactly what we do in our approach. Therefore, $\mathcal{T}_{\text{hard}}$ contains only hard formulas that appear in $\mathcal{T}$. Next, on lines 5 through 12, we check if the entire unmodified $\mathcal{T}_{\text{hard}}$ is compatible (since for $n = 0$, we do not remove any formulas). If it is compatible, we return an empty set indicating that all the hard formulas in the original seed theory $\mathcal{T}$ are compatible with the examples. If we detect incompatibility, we will need to remove some, and perhaps even all, hard formulas in order to arrive at a logically consistent theory. Therefore, we incrementally start removing $n$-tuples of formulas. That is, in the subsequent $|\mathcal{T}_{\text{hard}}|$ iterations of the while loop, we determine if we can restore consistency by removing any one of the hard formulas in $\mathcal{T}_{hard}$. If we can, we return the set $\mathcal{T}_{hard} \setminus f_i$, where $f_i$ is the identified and removed incompatible formula. If consistency cannot be restored by removing a single formula, we in turn begin considering pairs of formulas ($n = 2$), triples ($n = 3$), etc. until we find a pruned theory $\mathcal{T}^c$ that is consistent with *all* examples.

In general, we do need to consider $n$-tuples of formulas, rather than testing each formula in isolation. This is due to disjunctive formulas in conjunction with an possibly incomplete truth assignment in the training data. Consider the following theory in propositional logic:

$$f_1 = \neg a \vee b$$
$$f_2 = \neg b \vee c$$
$$\text{Data: } a \wedge \neg c$$

(Following the closed-world assumption, the negated atom $c$ would actually not appear in the training data, but we explicitly include it in this example for clarity.) While $f_1$ and $f_2$ are each individually consistent with the data, $f_1 \wedge f_2$ is inconsistent with the data. More complicated examples can be constructed, where every group of $k$ formulas is inconsistent with the data, even though the individual formulas are. In a special case where the truth values of all atoms in the training examples are known, the formulas can be tested for consistency individually, which reduces the original exponential number of iterations Algorithm 2 executes, in the worst case, to a linear complexity. An interesting direction for future work is to explore applications of logical methods to lower the computational cost for the general case of partially observed data.

We also note that some hard formulas model physical constraints or inviolable rules of capture the flag, and therefore hold *universally*. Appropriately, these formulas are not eliminated by Algorithm 2. As an example, consider formula H1 in Figure 5, which asserts that each player occupies exactly one cell at any given time. This formula is satisfied in games that include both successful and failed activities. On the other hand, consider formula H8 in the same figure. It contains a captured player to the cell he was captured in (following the "captured players cannot move" rule of CTF). While this holds for successful capturing events, it does not necessarily hold for failed attempts at capturing. Therefore, when rule H8 is expanded via second-order ML, only some of the derived formulas are going to be consistent with the observations.

Specifically, the candidate formula in Equation 6 will be pruned away, as it is inconsistent with the training examples, i.e., players that were only nearly captured continue to be free to move about. However, the remaining three variants of formula H8 will not be pruned away. Equation 7 will always evaluate to true, since if someone attempts to re-capture an already captured player $a$, $a$ does indeed remain stationary. Similarly, Equation 8 is also consistent with all the example CTF games because if there is a failed attempt at capture immediately followed by a successful capture, the captured player does remain in place from time $t$ onward. Finally, Equation 9 is compatible as well, since it is the original formula H8 that is consistent with the observations.

$$\forall a, t, c : \big( \text{isFailedCaptured}(a, t) \wedge \text{isFailedCaptured}(a, t + 1) \wedge \text{snap}(a, c, t) \big) \Rightarrow \text{snap}(a, c, t + 1) \tag{6}$$

$$\forall a, t, c : \big( \text{isCaptured}(a, t) \wedge \text{isFailedCaptured}(a, t + 1) \wedge \text{snap}(a, c, t) \big) \Rightarrow \text{snap}(a, c, t + 1) \tag{7}$$

$$\forall a, t, c : \big(\text{isFailedCaptured}(a, t) \wedge \text{isCaptured}(a, t+1) \wedge \text{snap}(a, c, t)\big) \Rightarrow \text{snap}(a, c, t+1) \quad (8)$$

$$\forall a, t, c : \big(\text{isCaptured}(a, t) \wedge \text{isCaptured}(a, t+1) \wedge \text{snap}(a, c, t)\big) \Rightarrow \text{snap}(a, c, t+1) \quad (9)$$

The function *testSAT()* (line 11 in Algorithm 2) checks whether a given candidate theory $\mathcal{T}^c$ is compatible with the examples $F$ by the following process. First, we ground $\mathcal{T}^c$ using the objects in $O$, thereby creating a ground theory $\mathcal{G}$. For example, if $\mathcal{T}^c = \{p(x) \Rightarrow q(x)\}$ and $O = \{B, W\}$, the grounding would be $\mathcal{G} = \{p(B) \Rightarrow q(B), p(W) \Rightarrow q(W)\}$. Then we check if $\mathcal{G} \cup F_{\text{hidden}}$ is satisfiable using the miniSAT solver, where $F_{\text{hidden}}$ is simply the set of hidden atoms that appear in $F$. Intuitively, this corresponds to testing whether we can "plug in" the worlds in $F$ into $\mathcal{T}^c$ while satisfying all the hard constraints. Though satisfiability is an NP-complete problem, in practice *testSAT()* completes within tenths of a second even for the largest problems in our CTF domain.

For instance, suppose $F_{\text{hidden}} = \{p(B), \neg q(B)\}$. Then we test satisfiability of the formula

$$\Big(p(B) \Rightarrow q(B)\Big) \wedge \Big(p(W) \Rightarrow q(W)\Big) \wedge p(B) \wedge \neg q(B).$$

In this case we cannot satisfy it since we are forced to set $p(B)$ to *true* and $q(B)$ to *false*, which renders the first clause—and therefore the whole formula—*false*.

An alternative approach to pruning formulas via satisfiability testing, as we have just described, would be to treat both types of formulas (hard and soft) in the inflated theory $\mathcal{M}'_S$ as strictly soft formulas and learning a weight for each formula from examples of both successful and failed game events. However, this introduces several complications that negatively impact the system's performance as well as model clarity. First, the number of formulas in the inflated theory can be exponentially larger than in the seed theory. While the instantiation of the second-order ML representation can be quantified to limit this expansion, we still have worst-case exponential blow-up. By treating all formulas as soft ones, we now need to potentially learn many more weights. This is especially problematic for activities that occur rarely, as we may not have enough training data to properly learn those weights. Eliminating the hard candidate formulas by proving them inconsistent dramatically reduces the number of parameters we have to model. While satisfiability testing is NP-complete, weight learning in Markov logic entails running inference multiple times, which is itself a #P-complete problem.

The second reason for distinguishing between soft and hard formulas is the resulting clarity and elegance of the final learned model $\mathcal{M}_{S+F}$. Even in situations when we have enough training data to properly learn a large number of weights, we run into overfitting problems, where neither the structure nor the parameters of the model represent the domain in a natural way. Our experiments have shown that if we skip the pruning stage (steps 3 and 4 in Algorithm 1), the model's recognition performance does not differ from that of a pruned model in a significant way (p-value of 0.45). However, we end up with a large number of soft formulas with a mixture of positive and negative weights that the learning algorithm carefully tuned and balanced to fit the training data. They however bear little relationship to the concepts in the underlying domain. Not only does this make it very hard for a human expert to analyze the model, but it makes it even harder to modify the model.

For these reasons, softening all hard formulas is, in general, infeasible. An interesting direction of future work will be to identify a small amount of *key* inconsistent hard formulas to soften, while eliminating the rest of the inconsistent hard formulas. This however entails searching in a large space of candidate subsets of softened formulas, where each iteration requires expensive re-learning of all weights.

Note that Algorithm 2 terminates as soon as it finds a compatible theory that requires the smallest number of formula-removals. We also experimented with an active learning component to our system, where we modify Algorithms 1 and 2 such that they present *several* possible refinements of the theory to the user who then selects the one that looks best. The proposed modifications are shown both at the ML theory level with modified sections (formulas) highlighted as well as at the data level where the program shows the inferred consequences of those modifications. For each candidate modification, the corresponding consequences are displayed as a collection of animations where each animation shows what the results of activity recognition would be if we committed to that particular candidate theory. Note that even people who do not have background in ML can interact with such a system since the visualization is easy to understand. Interestingly, in the case of captures and frees, the least modified theory that the "off-line" version of the algorithm finds is also the best one and therefore there is no need to query the user. One can view this as a differential variant of Occam's razor. However, for different activities or other domains, the active learning approach may be worth revisiting and we leave its exploration for future work.

Finally, general structure learning techniques from statistical-relational AI and from inductive logic programming are not applicable as a substitute for our theory augmentation algorithm for several reasons. The main reason is that, for efficiency reasons, existing techniques in the literature typically operate over a very restricted set of formula templates. That is, they consider only Horn clauses, or only formulas without an existential quantifier, or only formulas with at most $k$ literals or with at most $l$ variables, and so on. This set of restrictions is part of the language bias of any given approach. While in principle, structure learning is possible without a language bias, one often has to carefully define one for the sake of tractability (see the Section 7 for details). In our approach, the language bias is defined implicitly as discussed in Section 5.2.1.

### 5.3 Extracting The Goal From Success and Failure

Recall that applying the theory augmentation process (Algorithm 1) on the CTF seed theory of successful interactions (shown in Figures 5 and 6) induces a new set of formulas that capture the structure of failed activities and ties them together with the existing formulas in the seed theory.

The logically inconsistent formulas $\mathcal{I}$ that Algorithm 2 returns are ones that are not satisfiable in the worlds with failed activities. At the same time, variants of those formulas were consistent with the examples of successful actions occurring in the games. Therefore, $\mathcal{I}$ represents the *difference* between a theory that models only successful activities and the augmented theory of both successful and failed actions, that has been derived from it. Intuitively, the difference between success and failure can be viewed as the intended purpose of any given activity a rational agent executes, and consequently as the goal the agent has in mind when he engages in that particular activity. In the next section, we will explore the goals extracted from the CTF domain in this fashion.

This concludes discussion of our models and methodology, and now we turn to experimental evaluation of the framework presented above.

## 6. Experiments and Results

We evaluate our approach along the three major directions outlined in Section 5 (Methodology), while focusing on answering the four research questions formulated *ibidem*. The structure of this section closely follows that of the Methodology section.

In a nutshell, we are first interested in how our Markov logic models perform on the standard multi-agent activity recognition task—labeling successful activities—and how their performance compares to the alternative models. Second, we examine the augmented model that captures both successful and failed attempts at activities. This is the model $\mathcal{M}_{S+F}$ induced by Algorithm 1, which also lets us extract the intended goal of the activities in question. Third, we compare the performance of $\mathcal{M}_{S+F}$ on the task of jointly recognizing *all four* activities with that of an alternative model. Finally, we investigate to what extent the reasoning about failed attempts does help in recognition of successfully executed activities.

All experiments are performed on our capture the flag dataset consisting of four separate games. The dataset is summarized in Table 4, where for each game we list the number of raw GPS readings and the number of instances of each activity of interest. We evaluate the models via four-fold cross-validation, always training on three games (if training is required for a model) and testing against the fourth. For each experimental condition below, we report precision, recall, and F1 scores attained by each respective model over the four cross-validation runs. We have purposefully chosen to split the data so that each cross-validation fold directly corresponds to a separate game of CTF for conceptual convenience and clarity. As we discussed above, the events occurring in the games often have far-reaching consequences. For example, most captured players are never freed by their allies. Therefore, a capture at the beginning of a game typically profoundly influences the entire rest of the game. For this reason, splitting the games randomly or even manually would introduce unnecessary complications, as most of the segments would have dependencies on other segments. By enforcing that each fold exactly corresponds with a different game, we make each fold self-contained.

To quantify the statistical significance of the pair-wise differences between models, we use a generalized probabilistic interpretation of F1 score (Goutte & Gaussier, 2005). Namely, we express F1 scores in terms of gamma variates derived from models' true positives, false positives, and false negatives ($\lambda = 0.5$, $h = 1.0$, *cf.*, Goutte & Gaussier, 2005). This approach makes it possible to compare our results to future work that may apply alternative models on similar, but not identical, datasets. A future comparison may, for instance, include additional games or introduce random splits of the data. We note that standard statistical significance tests cannot be applied in those situations. All p-values reported are one sided, as we are interested if models' performance significantly *improves* as their level of sophistication increases.

### 6.1 Recognition of Successful Activities

Recall that for both our two-step (2SML) and unified (UML) Markov logic models, we specify the Markov logic formulas by hand and optimize the weights of the soft formulas via supervised online learning. We run a modified version of theBeast software package to perform weight learning and MAP inference. theBeast implements the cutting plane meta solving scheme for inference in Markov logic, where the ground ML network is reduced to an integer linear program that is subsequently solved by the LpSolve ILP solver. We chose this approach as opposed to, e.g., MaxWalkSAT that can get "stuck" at a local optimum, since the resulting run times are still relatively short (under an hour even for training and testing even the most complex model).

| | #GPS | #AC | #FC | #AF | #FF |
|---|---|---|---|---|---|
| **Game 1** | 13,412 | 2 | 15 | 2 | 1 |
| **Game 2** | 14,420 | 2 | 34 | 2 | 1 |
| **Game 3** | 3,472 | 6 | 12 | 0 | 2 |
| **Game 4** | 10,850 | 3 | 4 | 1 | 0 |
| **Total** | 42,154 | 13 | 65 | 5 | 4 |

Table 4: CTF dataset overview: #GPS is the total number of raw GPS readings, #AC and #FC is the number actual (successful) and failed captures respectively, and analogously for freeings (#AF and #FF).

At weight learning time, we use the margin infused relaxed algorithm (MIRA) for weight updates while the loss function is computed from the number of false positives and false negatives over the hidden atoms, as described in the Methodology section. The discretization step for the dynamic Bayesian network model (DBN) is implemented in Markov logic and is also executed in this fashion. The DBN model is trained via maximum likelihood as described in Section 4.2. The two deterministic baselines (B and B+S) do not require any training phase.

At inference time, we are interested in the most likely explanation of the data. In Markov logic, maximum *a posteriori* inference reduces to finding a complete truth assignment that satisfies all the hard constraints while maximizing the sum of the weights of the satisfied soft formulas. At testing time, theBeast Markov logic solver finds the most likely truth assignment to the hidden atoms as described above, and in this section we are specifically interested in the values of the capturing and freeing atoms.

In DBNs, the most likely explanation of the observations is equivalent to Viterbi decoding. The DBN model assigns either free or captured state to each player for every time step. We then label all transitions from free to captured state as capturing and all transitions from captured to free as freeing. Note that the DBN model is capable of determining which player is being freed or captured, but it does not model which player does the freeing or capturing. In our evaluation, we give it the benefit of the doubt and assume it always outputs the correct actor.

For all models, inference is done simultaneously over an entire game (on average, about 10 minutes worth of data). Note that we do not restrict inference to a (small) sliding time window. As the experiments described below show, many events in this domain can only be definitely recognized long after they occur. For example, GPS noise may make it impossible to determine whether a player has been captured at the moment of encounter with an enemy, but as the player thereafter remains in place for a long time, the possibility of his capture becomes certain.

Figures 7 and 8 summarize the performance of our models of successful capturing and freeing in terms of precision, recall, and F1 score calculated over the four cross-validation runs. For clarity, we present the results in two separate plots, but each model was jointly labeling both capturing and freeing activities. We do not consider the baseline model for freeing recognition as that activity makes little sense without having a notion of player state (captured or free).

We see that the unified approach yields the best results for both activities. Let us focus on capturing first (Figure 7). Overall, the unified model labels 11 out of 13 captures correctly—there

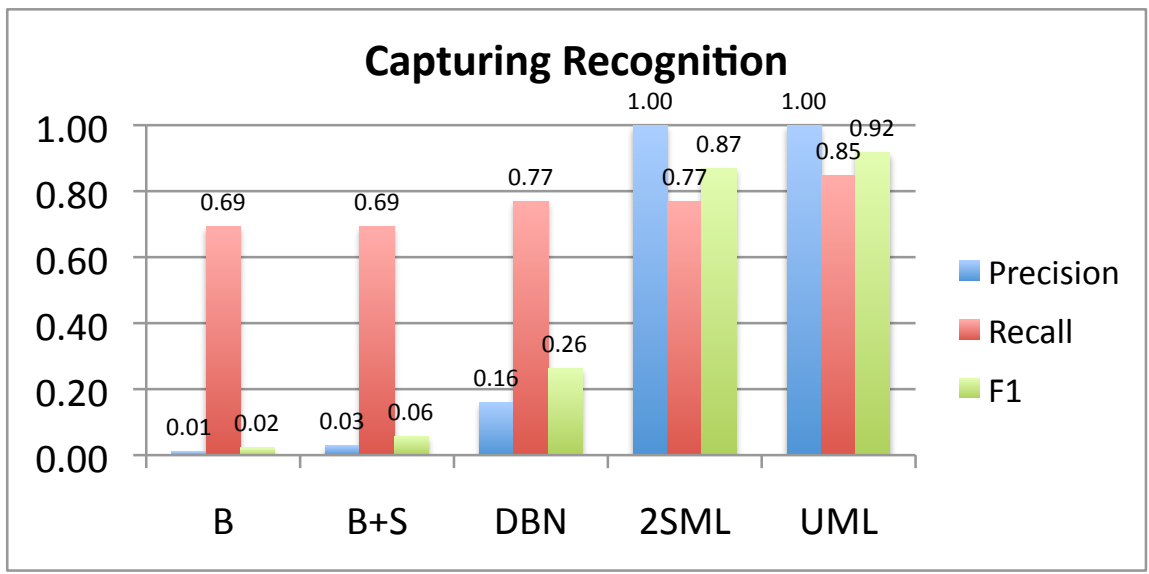

Figure 7: Comparison of performance of the five models on capturing recognition while doing joint inference over both capturing and freeing events. See Table 5 for statistical significance analysis of the pairwise differences between models. (B = baseline model, B+S = baseline model with states, 2SML = two-step Markov logic model, UML = unified Markov logic model)

are only two false negatives. In fact, these two capture events are missed by *all* the models because they involve two enemies that appear unusually far apart (about 12 meters) in the raw data. Even the unified approach fails on this instance since the cost of adjusting the players' trajectories—thereby losing score due to violation of the geometry-based constraints—is not compensated for by the potential gain from labeling an additional capture.

Note that even the two-step approach recognizes 10 out of 13 captures. As compared to the unified model, it misses one additional instance in which the involved players, being moderately far apart, are snapped to mutually nonadjacent cells. On the other hand, the unified model does not fail in this situation because it is not limited by prior nonrelational snapping to a few nearby cells. However, the difference between their performance on our dataset is not statistically significant even at the 0.05 level (p-value of 0.32).

Both deterministic baseline models (B and B+S) perform very poorly. Although they yield a respectable recall, they produce an overwhelming amount of false positives. This shows that even relatively comprehensive pattern matching does not work at all in this domain. Interestingly, the performance of the DBN model leaves much to be desired as well, especially in terms of precision. While the DBN model is significantly better than both baselines (p-value less than $5.9 \times 10^{-5}$), it also achieves significantly worse performance than both the Markov logic models (p-value less than 0.0002; see Table 5).

Table 5 summarizes p-values of pairwise differences between models of actual (i.e., successful) capturing. While the difference between the Markov logic-based models (2SML and UML) are not statistically significant (p-value of 0.32), pairwise differences in F1 scores between all other models are significant at the 0.02 level, and most often even at much lower p-values.

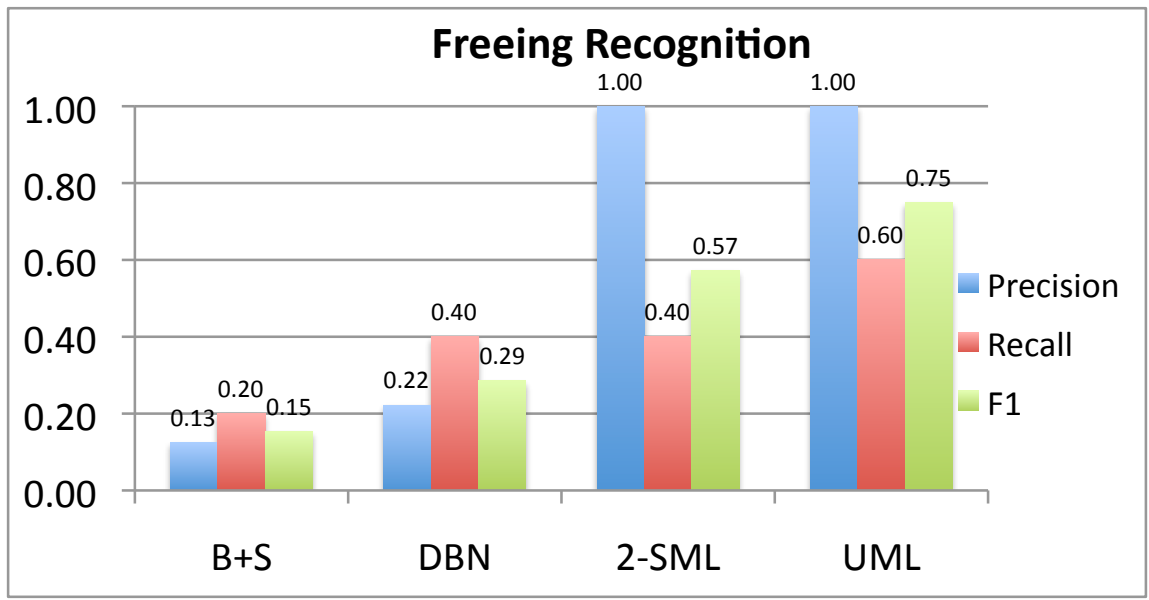

Figure 8: Comparison of performance of our three models on freeing recognition while doing joint inference over both capturing and freeing events. See Table 6 for statistical significance analysis of the pairwise differences between models. (B+S = baseline model with states, 2SML = two-step Markov logic model, UML = unified Markov logic model)

| | B+S | DBN | 2SML | UML |
|---|---|---|---|---|
| **B** | 0.0192 | $3.6 \times 10^{-6}$ | $5.1 \times 10^{-7}$ | $2.9 \times 10^{-7}$ |
| **B+S** | - | $5.9 \times 10^{-5}$ | $9.4 \times 10^{-6}$ | $1.4 \times 10^{-6}$ |
| **DBN** | - | - | 0.0002 | $8.0 \times 10^{-5}$ |
| **2SML** | - | - | - | 0.3230 |

Table 5: Summary of statistical significance (one sided p-values) of the pairwise differences between F1 scores for models of actual capturing. (B = baseline model, B+S = baseline model with states, DBN = dynamic Bayesian network model, 2SML = two-step Markov logic model, UML = unified Markov logic model)

Though the unified model still outperforms its alternatives in the case of freeing recognition as well, its performance is further from ideal as compared to the capture recognition case (Figure 8). It correctly identifies only 3 out of 5 freeing events in the games, but does not produce any false positives. This is partly due to the dependency of freeing on capturing. A failure of a model to recognize a capture precludes its recognition of a future freeing. Another reason is the extreme sparseness of the freeing events (there are only five of them in 40,000+ datapoints). Finally, in some instances players barely move after they had been freed. This may occur for a number of reasons ranging from already occupying a strategic spot to simply being tired. Such freeing instances are very challenging for any automated system, and even people familiar with the game to recognize (several situations would have been extremely hard to disambiguate if we didn't have access to our notes about data collection).

|        | DBN    | 2SML   | UML    |
|--------|--------|--------|--------|
| **B+S**  | 0.2739 | 0.0733 | 0.0162 |
| **DBN**  | -      | 0.1672 | 0.0497 |
| **2SML** | -      | -      | 0.2743 |

Table 6: Summary of statistical significance (one sided p-values) of the pairwise differences between F1 scores for models of actual freeing. (B+S = baseline model with states, DBN = dynamic Bayesian network model, 2SML = two-step Markov logic model, UML = unified Markov logic model)

The two-step ML model does a slightly worse job than the unified model on freeing recognition. It correctly identifies only 2 out of 5 freeings for the same reasons as in the capturing recognition case. Similarly to models of actual captures, the difference between the unified and two-step freeing models is not statistically significant (p-value of 0.27).

Table 6 summarizes p-values of pairwise differences between models of actual (i.e., successful) freeing. Here we see that only the difference between B+S and UML models is statistically significant (p-value of 0.01), whereas the differences between the rest of the model pairs are not statistically significant. Since there are only five instances of successful freeing, the 2SML model does not perform significantly better than the B+S model at the 0.05 significance level (p-value of 0.07). However, the UML model achieves better recognition results than even the DBN model with high confidence (p-value less than 0.05). Therefore, we see that although the 2SML model strictly dominates the non-Markov logic models when evaluated on capturing recognition, we need the full power of the unified ML model to strictly outperform the nonrelational alternatives for freeing. This suggests that as we move to more complex and more interdependent activities, relational and unified modeling approaches will be winning by larger and larger margins.

Even though the statistical significance tests suggest that 2SML is likely to give similar results to UML, it is important to note that 2SML, by design, *precludes* recognition of the activities in question in certain situations. Namely, as our experiments demonstrate, when the players are snapped to cells that are too far apart, the two-step model does not even consider those instances as candidates for labeling, and inevitably fails at recognizing them. Therefore, one needs to look beyond the p-values obtained when comparing the fully unified models to various alternatives.

As expected from the experiments with capturing recognition, both deterministic baseline models perform very poorly on freeing recognition as well. Not only do they produce an overwhelming amount of false positives, they also fail to recognize most of the freeing events.

Thus, we see that the models cast in Markov logic perform significantly better than both of the deterministic baseline models, and also better than the probabilistic, but nonrelational, DBN model. We note that the DBN model has the potential to be quite powerful and similar DBNs have been applied with great success in previous work on activity recognition from location data (Eagle &

Pentland, 2006; Liao, Patterson, Fox, & Kautz, 2007). It also has many similarities with the two-step ML model. They both share the same denoising and discretization step, and they both operate on the same observed data. The key difference is that the DBN model considers players individually, whereas the two-step ML model performs joint reasoning.

Looking at the actual CTF game data, we see several concrete examples of how this hurts DBN's labeling accuracy. For instance, consider a situation where two allies had been captured near each other. Performing inference about individual players in isolation allows the DBN model to infer that the two players effectively free each other, even though in reality they are both captured and cannot do so. This occurs because the DBN model is oblivious to the explicit states of one's teammates as well as opponents. Since capturing and freeing are interdependent, the obliviousness of the DBN model to the state of the actors negatively impacts its recognition performance for both activities. The example we just gave illustrates one type of freeing false positives. The hallucinated freeings create opportunities that often lead to false positives of captures, creating a vicious cycle. False negatives of freeing (capturing) events often occur for players who the model incorrectly believes have already been freed (captured) at a prior time.

Since the Markov logic based models are significantly better—with a high level of confidence—than the alternatives that are not fully relational, the experiments above validate our hypothesis that we need to exploit the rich relational and temporal structure of the domain in a probabilistic way and at the same time affirmatively answer research question Q1 (*Can we reliably recognize complex multi-agent activities in the CTF dataset even in the presence of severe noise?*). Namely, we show that although relatively powerful probabilistic models are not sufficient to achieve high labeling accuracy, we can gain significant improvements by formulating the recognition problem as learning and inference in Markov logic networks.

Now we turn to the evaluation of our method of learning models of both success and failure in people's activities.

## 6.2 Learned Formulas and Intentions

Applying the theory augmentation process (Algorithm 1) on the CTF seed theory (shown in Figures 5 and 6) induces a new set of formulas that capture the structure of failed activities and ties them together with the existing formulas in the theory. We call this model $\mathcal{M}_{S+F}$. Figure 9 shows examples of new weighted formulas modeling failed freeing and capturing attempts that appear in $\mathcal{M}_{S+F}$.

First, note that our system correctly carries over the basic preconditions of each activity (contrast formulas S4 with S4′ and S5 with S5′ in Figures 6 and 9 respectively). This allows it to *reliably* recognize both successful and failed actions instead of, e.g., merely labeling all events that at some point in time appear to resemble a capture as near-capture. This re-use of preconditions directly follows from the language bias of the theory augmentation algorithm.

Turning our attention to the learned hard formulas, we observe that the system correctly induced equivalence classes of the states, and also derived their mutual exclusion relationships (H5′). It furthermore tied the new failure states to their corresponding instantaneous interactions (H6′ and H7′).

Finally, the algorithm correctly discovers that the rule *"If a player is captured then he or she must remain in the same location"* (H8, Figure 5) is the key distinction between a successful and failed capture (since players who were not actually captured can still move). Therefore, it introduces an appropriate rule for the failed captures (H8′, Figure 9) explicitly stating that failed capturing does not confine the near-captured player to remain in stationary. An analogous process yields a fitting separation between failed and successful freeings. Namely, our model learns that an unsuccessfully freed player remains stationary. This learned difference between success and failure in players' actions directly corresponds to the goal of the activity and consequently the intent of rational actors. This difference is what our system outputs as the intended goal of capturing activity (and analogously for freeing).

These experimental results provide an evidence for a resounding "yes" to both Q2 (*Can models of attempted activities be automatically learned by leveraging existing models of successfully performed actions?*) and Q3 (*Does modeling both success and failure allow us to infer the respective goals of the activities?*) within the CTF domain.

We note that instead of applying our automated theory augmentation method, a person could, in principle, manually formulate a Markov logic theory of successful as well as failed activities by observing the games. After all, this is how we designed the initial seed model of successful events. However, this process is extremely time consuming, as one tends to omit encoding facts that to us, humans, seem self-evident but need to be explicitly articulated for the machine (e.g., a single person cannot be at ten different places at once, or that a player is either free or captured but not both). It is also surprisingly easy to introduce errors in the theory, that are difficult to debug, mostly because of the complex weight learning techniques involved. Therefore, we believe that the theory augmentation method is a significant step forward in enhancing models' capabilities while requiring small amounts of human effort. As the complexity of domains and their models increases, this advantage will gain larger and larger importance.

### 6.3 Recognition of Both Successful and Failed Activities

We now compare the performance of our model $\mathcal{M}_{S+F}$ to an alternative (baseline) method that labels all four activities in the following way. Similarly to the baseline with states model for successful interactions defined in Section 5.1, there are two separate stages. First we snap each GPS reading to the nearest cell by applying only the geometric constraints (H1 and S1–S3) of our theory, and afterward we label the instances of our activities. The following labeling rule is applied. We loop over the whole discretized (via snapping) data set and look for instances where a pair of players $a$ and $b$ were snapped (in the first step) to either the same cell or to two adjacent cells at time $t$, they are enemies, $b$ is not captured already, and $a$ is on its home territory while $b$ is not. If $b$ moves (is snapped to a different cell at a later time) *without* having an ally nearby, we output *failedCapturing(a,b,t)*, otherwise we output *capturing(a,b,t)*. The labeling rule for freeing is defined analogously and all four events are tied together. We also tested a variant of the DBN model introduced in Section 5.1 that has two additional hidden state values for node $S_t$: isFailedFree and isFailedCaptured. However, the difference in the results obtained with this model was not statistically significant (p-value of 0.38), and therefore we focus on the conceptually more straightforward baseline model described above.

Model $\mathcal{M}_{S+F}$ is evaluated using four-fold cross-validation (always training on three games and testing against the fourth). Figure 10 compares both models in terms of precision, recall, and F1 score. Note that all four activities are modeled *jointly* in both models. The F1 score of the augmented model is significantly better than that of the baseline for all four target activities (p-value less than $1.3 \times 10^{-4}$).

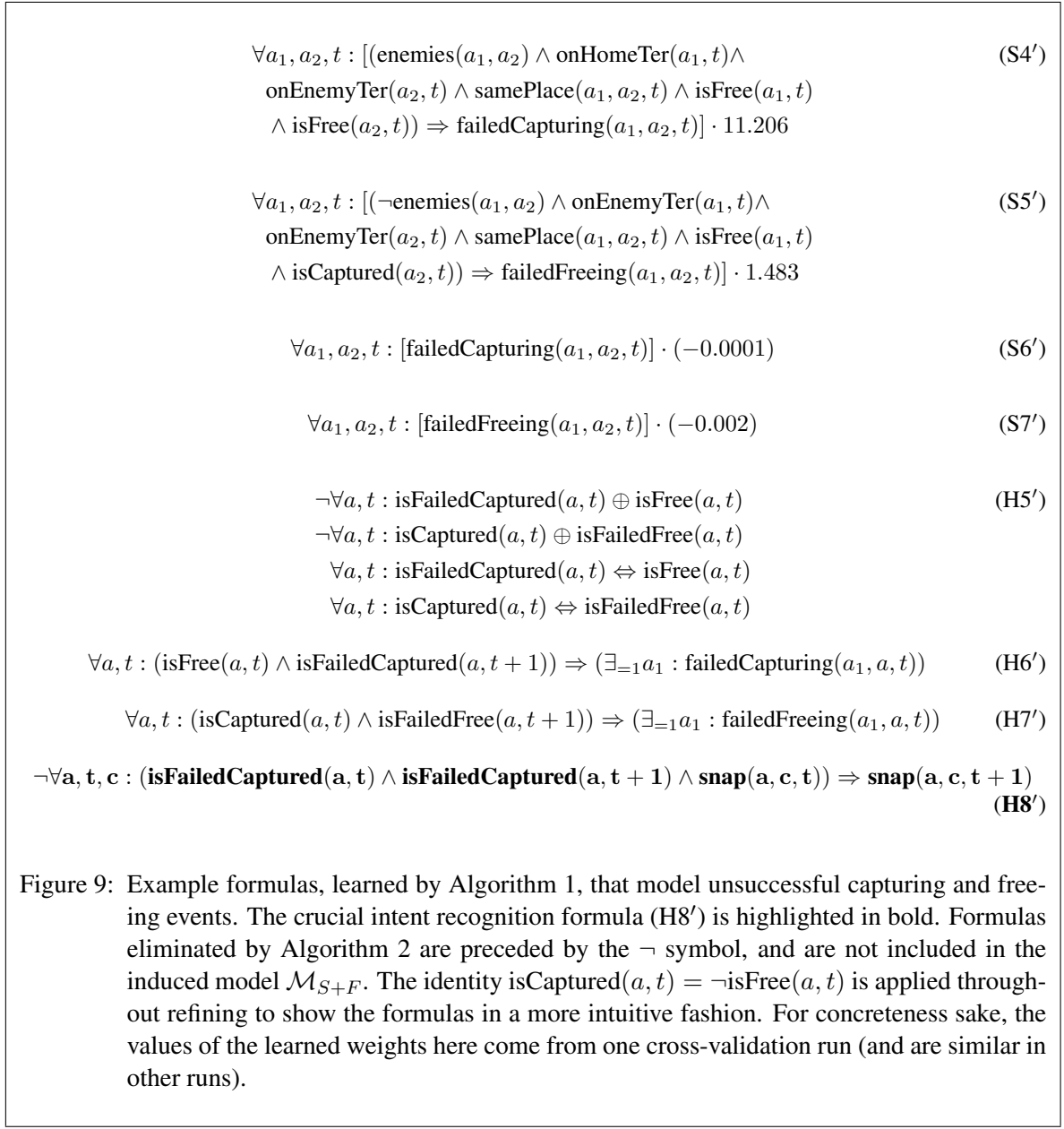

$$\forall a_1, a_2, t : [(\text{enemies}(a_1, a_2) \wedge \text{onHomeTer}(a_1, t) \wedge \qquad\qquad\qquad (\text{S4}')$$
$$\text{onEnemyTer}(a_2, t) \wedge \text{samePlace}(a_1, a_2, t) \wedge \text{isFree}(a_1, t)$$
$$\wedge \, \text{isFree}(a_2, t)) \Rightarrow \text{failedCapturing}(a_1, a_2, t)] \cdot 11.206$$

$$\forall a_1, a_2, t : [(\neg\text{enemies}(a_1, a_2) \wedge \text{onEnemyTer}(a_1, t) \wedge \qquad\qquad\qquad (\text{S5}')$$
$$\text{onEnemyTer}(a_2, t) \wedge \text{samePlace}(a_1, a_2, t) \wedge \text{isFree}(a_1, t)$$
$$\wedge \, \text{isCaptured}(a_2, t)) \Rightarrow \text{failedFreeing}(a_1, a_2, t)] \cdot 1.483$$

$$\forall a_1, a_2, t : [\text{failedCapturing}(a_1, a_2, t)] \cdot (-0.0001) \qquad\qquad\qquad (\text{S6}')$$

$$\forall a_1, a_2, t : [\text{failedFreeing}(a_1, a_2, t)] \cdot (-0.002) \qquad\qquad\qquad (\text{S7}')$$

$$\neg\forall a, t : \text{isFailedCaptured}(a, t) \oplus \text{isFree}(a, t) \qquad\qquad\qquad (\text{H5}')$$
$$\neg\forall a, t : \text{isCaptured}(a, t) \oplus \text{isFailedFree}(a, t)$$
$$\forall a, t : \text{isFailedCaptured}(a, t) \Leftrightarrow \text{isFree}(a, t)$$
$$\forall a, t : \text{isCaptured}(a, t) \Leftrightarrow \text{isFailedFree}(a, t)$$

$$\forall a, t : (\text{isFree}(a, t) \wedge \text{isFailedCaptured}(a, t+1)) \Rightarrow (\exists_{=1}a_1 : \text{failedCapturing}(a_1, a, t)) \quad (\text{H6}')$$

$$\forall a, t : (\text{isCaptured}(a, t) \wedge \text{isFailedFree}(a, t+1)) \Rightarrow (\exists_{=1}a_1 : \text{failedFreeing}(a_1, a, t)) \quad (\text{H7}')$$

$$\boldsymbol{\neg\forall a, t, c : (\text{isFailedCaptured}(a, t) \wedge \text{isFailedCaptured}(a, t+1) \wedge \text{snap}(a, c, t)) \Rightarrow \text{snap}(a, c, t+1)}$$
$$\boldsymbol{(\text{H8}')}$$

Figure 9: Example formulas, learned by Algorithm 1, that model unsuccessful capturing and freeing events. The crucial intent recognition formula (H8') is highlighted in bold. Formulas eliminated by Algorithm 2 are preceded by the ¬ symbol, and are not included in the induced model $\mathcal{M}_{S+F}$. The identity isCaptured$(a, t) = \neg$isFree$(a, t)$ is applied throughout refining to show the formulas in a more intuitive fashion. For concreteness sake, the values of the learned weights here come from one cross-validation run (and are similar in other runs).

We see that the baseline model has, in general, a respectable recall but it produces a large number of false positives for all activities. The false positives stem from the fact that the algorithm is "greedy" in that it typically labels a situation where several players appear close to each other for certain period of time as a sequence of many captures and subsequent frees even though none of them actually occurred. Model $\mathcal{M}_{S+F}$ gives significantly better results because it takes full advantage of the structure of the game in a probabilistic fashion. It has a similar "over labeling" tendency only in the case of failed captures, where a single capture attempt is often labeled as several consecutive attempts. While this hurts the precision score, it is not a significant deficiency, as in practice, having a small number of short game segments labeled as possible near-captures is useful as well.

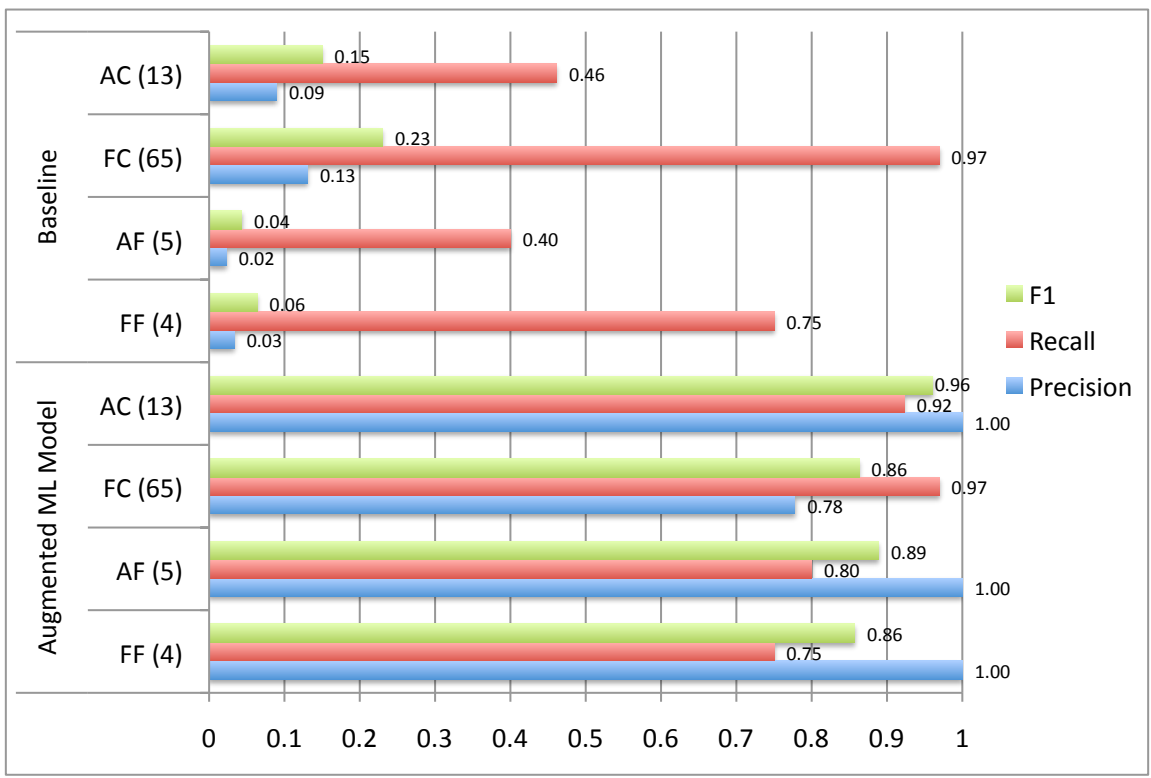

Figure 10: Performance of the baseline and augmented ($\mathcal{M}_{S+F}$) models on joint recognition of successful and failed capturing and freeing. The F1 score of the augmented model is significantly better than that of the baseline for all four target activities (p-value less than $1.3 \times 10^{-4}$). AC = actual (successful) capturing, FC = failed capturing, AF = actual freeing, FF = failed freeing.

We also note that even though the original model (UML) did not contain any information on failed capturing nor failed freeing, the performance of $\mathcal{M}_{S+F}$ is respectable even for those two newly introduced activities. We only provided examples of game situations where those attempts occur and the system augmented itself and subsequently labeled all four activities. Thus, we see that we can indeed extend preexisting models in an automated fashion so that the unified model is capable of recognizing not only individual activities, but also both success and failure in people's behavior.

## 6.4 The Effect of Modeling Failed Attempts on Recognition of Successful Activities

To address research question Q4 (*Does modeling failed attempts of activities improve the performance on recognizing the activities themselves?*), we want to see how much does the recognition of attempted activities help in modeling the successful actions (the latter being the standard activity recognition problem). Toward that end, we compare the Markov logic model $\mathcal{M}_S$ that jointly labels only successful capturing and freeing with model $\mathcal{M}_{S+F}$ that jointly labels both successful and failed attempts at both capturing and freeing (see Section 5.2.1 for a detailed description of the two models). However, we evaluate them in terms of precision, recall, and F1 score only on *successful* interactions, not all four types of activities.

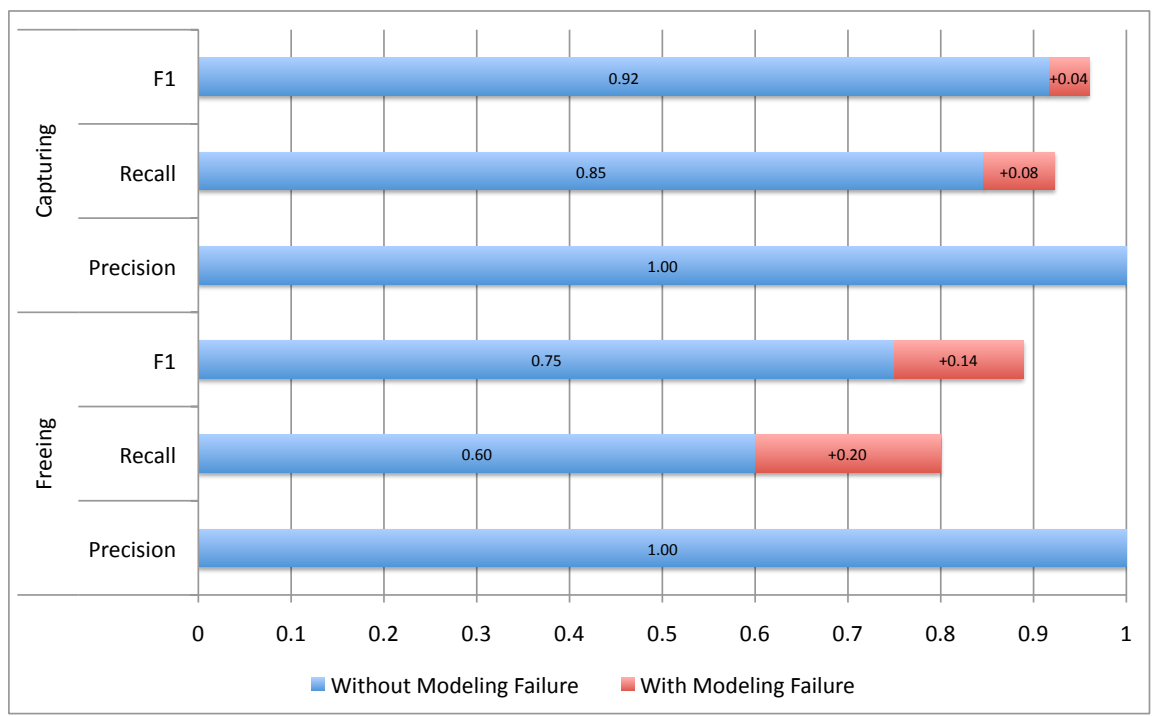

Figure 11: Considering unsuccessfully attempted activities strictly improves performance on standard activity recognition. Blue bars show scores obtained with the unified Markov logic model that considers only *successful* activities ($\mathcal{M}_S$). The red bars indicate the additive improvement provided by the augmented model that considers both *successful and failed* activities ($\mathcal{M}_{S+F}$, the output of Algorithm 1). Each model labels its target activities jointly, we separate capturing and freeing in the plot for clarity. Precision has value of 1 for both models. F1 scores obtained when explicitly modeling failed attempts are not statistically different from F1 scores obtained without modeling attempts at a high confidence level (p-value of 0.20). However, these results still show the importance of reasoning about people's attempts when recognizing their activities; see text for details.

Figure 11 summarizes the results. We see that when evaluated on actual capturing, $\mathcal{M}_{S+F}$ performs better than $\mathcal{M}_S$, and similarly for freeing. However, the difference in F1 scores between a model that captures both attempted and successful activities ($\mathcal{M}_{S+F}$) and a model of only successful activities ($\mathcal{M}_S$) is not statistically significant (p-value of 0.20). This is partly because $\mathcal{M}_S$ already produces very solid results, leaving little room for improvement. Additionally, the CTF dataset contains relatively few events of interest. In terms of labeling performance at testing time, the difference between the two models is more than 11% ($\mathcal{M}_S$ and $\mathcal{M}_{S+F}$ recognize, respectively, 14 and 16 out of 18 successful activities correctly). Thus, we believe the trends shown in Figure 11 are promising and modeling attempted actions does improve recognition performance on both capturing and freeing, but evaluation on a dataset with a larger number of events is needed to show the difference to be statistically significant at a higher confidence level. However, this does not mean that recognizing attempts is unimportant. As we show above, our induced augmented model does recognize failed (as well as successful) activities in the complex CTF domain with high accuracy, and we argue this to be a significant contribution.

Finally, the comparison of $\mathcal{M}_S$ and $\mathcal{M}_{S+F}$ shows that applying our learning algorithm that augments a model with more recognition capabilities *does not hurt* model labeling performance. The fact that binary classification problems are typically easier to solve than their multi-class counterparts has been well reported on in machine learning literature (Allwein, Schapire, & Singer, 2001). Therefore, introducing new activities into a model, especially in an automated way, is likely to degrade its performance. Contrary to this intuition, our experiments show that $\mathcal{M}_{S+F}$ is no worse than $\mathcal{M}_S$ on successful activity recognition (i.e., their intersection) with high confidence, even though $\mathcal{M}_{S+F}$ is clearly richer and more useful.

## 7. Related Work

In the world of *single-agent* location-based reasoning, the work of Bui (2003) presents and evaluates a system for probabilistic plan recognition cast as an abstract hidden Markov memory model. Subsequently, the work of Liao et al. (2004) implements a system for denoising raw GPS traces and simultaneously inferring individuals' mode of transportation (car, bus, etc.) and their goal destination. They cast the problem as learning and inference in a dynamic Bayesian network and achieve encouraging results. In a follow-up work, Liao et al. (2005) introduce a framework for location-based activity recognition, which is implemented as efficient learning and inference in a relational Markov network.

The work of Ashbrook and Starner (2003) focuses on inferring significant locations from raw GPS logs via clustering. The transition probabilities between important places are subsequently used for a number of user modeling tasks, including location prediction. The work of Eagle and Pentland (2006) explores harnessing data collected on regular smart phones for modeling human behavior. Specifically, they infer individuals' general location from nearby cell towers and Bluetooth devices at various times of day. Applying a hidden Markov model (HMM), they show that predicting if a person is at home, at work, or someplace else can be achieved with more than 90% accuracy. Similarly, the work of Eagle and Pentland (2009) extracts significant patterns and signatures in people's movement by applying eigenanalysis to smart phone logs.

The work of Hu, Pan, Zheng, Liu, and Yang (2008) concentrates on recognition of interleaving and overlapping activities. They show that publicly available academic datasets contain a significant number of instances of such activities, and formulate a conditional random field (CRF) model that is capable of detecting them with high (more than 80%) accuracy. However, they focus solely on single-agent household activities.

People's conversation has been the primary focus of *multi-agent* modeling effort (Barbuceanu & Fox, 1995). In the fields of multi-agent activity recognition and studies of human behavior, researchers have either modeled conversation explicitly (e.g., Busetta, Serafini, Singh, & Zini, 2001), or have leveraged people's communication implicitly via call and location logs from mobile phones. This data has been successfully used to infer social networks, user mobility patterns, model socially significant locations and their dynamics, and others (Eagle & Pentland, 2006; Eagle, Pentland, & Lazer, 2009). This is arguably an excellent stepping stone for full-fledged multi-agent activity recognition since location is, at times, practically synonymous with one's activity (e.g., being at a store often implies shopping) (Tang, Lin, Hong, Siewiorek, & Sadeh, 2010), and our social networks have tremendous influence on our behavior (Pentland, 2008).

Additionally, a number of researchers in machine vision have worked on the problem of recognizing events in videos of sporting events, such as impressive recent work on learning models of baseball plays (Gupta et al., 2009). Most work in that area has focused on recognizing individual actions (e.g., catching and throwing), and the state of the art is just beginning to consider relational actions (e.g., the ball is thrown from player A to player B). The computational challenges of dealing with video data make it necessary to limit the time windows of a few seconds. By contrast, we demonstrate in this work that many events in the capture the flag data can only be disambiguated by considering arbitrarily long temporal sequences. In general, however, both our work and that in machine vision rely upon similar probabilistic models, and there is already some evidence that statistical-relational techniques similar to Markov logic can be used for activity recognition from video (Biswas, Thrun, & Fujimura, 2007; Tran & Davis, 2008).

Looking beyond activity recognition, recent work on relational spacial reasoning includes an attempt to locate—using spacial abduction—caches of weapons in Iraq based on information about attacks in that area (Shakarian, Subrahmanian, & Spaino, 2009). Additionally, the work of Abowd et al. (1997) presents a location- and context-aware system, Cyberguide, that helps people explore and fully experience foreign locations. Other researchers explore an intelligent and nonintrusive navigation system that takes advantage of predictions of traffic conditions along with a model of user's knowledge and competence (Horvitz et al., 2005). Finally, the work of Kamar and Horvitz (2009) explore automatic generation of synergistic plans regarding sharing vehicles across multiple commuters.

An interesting line of work in cognitive science focuses on intent and goal recognition in a probabilistic framework (Baker, Tenenbaum, & Saxe, 2006, 2007). Specifically, they cast goal inference as inverse planning problem in Markov decision processes, where Bayesian inversion is used to estimate the posterior distribution over possible goals. Recent extensions of this work begin to consider simulated multi-agent domains (Baker, Goodman, & Tenenbaum, 2008; Ullman, Baker, Macindoe, Evans, Goodman, & Tenenbaum, 2010; Baker, Saxe, & Tenenbaum, 2011). Comparison of the computational models against human judgement in synthetic domains shows a strong correlation between people's predicted and actual behavior. However, the computational challenges involved in dealing with the underlying partially observable Markov decision processes are prohibitive in more complex domains with large state spaces, such as ours.

The focus of our work is on a different aspect of reasoning about people's goals. Rather than inferring a distribution over possible, *a priori* known goals, we automatically *induce* the goals of complex multi-agent activities themselves.

Other researchers have concentrated on modeling behavior of people and general agents as reinforcement learning problems in both single-agent and multi-agent settings. The work of Ma (2008) proposes a system for household activity recognition cast as a single-agent Markov decision process problem that is subsequently solved using a probabilistic model checker. Wilson and colleagues address the problem of learning agents' *roles* in a multi-agent domain derived from a real-time strategy computer game (Wilson, Fern, Ray, & Tadepalli, 2008; Wilson, Fern, & Tadepalli, 2010). Experiments in this synthetic domain show strongly encouraging results. While we do not perform role learning ourselves, we anticipate that the work of Wilson et al. is going to play an important role in learning hierarchies of people's activities. In our capture the flag domain, one can imagine automatically identifying a particular player as, for example, a defender and subsequently leveraging this information to model his or her behavior in a more "personalized" way.

The work of Hong (2001) concentrates on recognizing the goal of an agent in the course of her activities in a deterministic, but relational setting. Interesting work on goal recognition has been also applied to computer-aided monitoring of complex multi-agent systems, where relationships between agents are leveraged to compensate for noise and sparse data (Kaminka, Tambe, Pynadath, & Tambe, 2002). By contrast, in our work we focus on *learning* the respective goals of a given set of multi-agent activities in a probabilistic setting. The knowledge is in turn leveraged to achieve a stronger robustness of the other recognition tasks. Similarly to the approach of Hong, our system does not need a supplied plan library either.

Our work also touches on anomaly detection since our system reasons about the failed attempts of the players. Anomaly detection concerns itself with revealing segments of the data that in some way violate our expectations. For an excellent survey of the subject, we refer the reader to the results of Chandola, Banerjee, and Kumar (2009). In the realm of anomaly detection within people's activities, the work of Moore and Essa (2001) addresses the problem of error detection and recovery card games that involve two players recorded on video. Their system models the domain with a stochastic context-free grammar and achieves excellent results.

We note that recognizing a failed attempt at an activity is more fine-grained a problem than anomaly detection. The failed event is not just anomalous in general.[7] Rather, it is the specific distinction between success and failure in human activities that we are interested in. And the distinction lies in the fact that an unsuccessful attempt does not yield a certain desired state whereas a successful action does. This desired state is exactly what our approach extracts for each activity in question. To our knowledge, there exists no prior work on explicit modeling and recognition of attempted activities or on learning the intended purpose of an activity in a multi-agent setting.

One of the components of our contribution focuses on *joint* learning and inference across multiple tasks (capturing, freeing, and their respective attempted counterparts). This is in contrast with the traditional "pipeline" learning architecture, where a system is decomposed into a series of modules and each module performs partial computation and passes the result on to the next stage. The main benefits of this set-up are reduced computational complexity and often higher modularity. However, since each stage is myopic, it may not take full advantage of dependencies and broader patterns within the data. Additionally, even though errors introduced by each module may be small, they can accumulate beyond tolerable levels as data passes through the pipeline.

An extensive body of work has shown that joint reasoning improves model performance in a number of natural language processing and data mining tasks including information extraction (i.e., text segmentation coupled with entity resolution) (Poon & Domingos, 2007), co-reference resolution (Poon & Domingos, 2008), information extraction coupled with co-reference resolution (Wellner, McCallum, Peng, & Hay, 2004), temporal relation identification (Yoshikawa, Riedel, Asahara, & Matsumoto, 2009; Ling & Weld, 2010), and record de-duplication (Domingos, 2004; Culotta & McCallum, 2005). Similarly to our work, some of the above models are cast in Markov logic. However, prior work uses sampling techniques to perform learning and inference, whereas we apply a reduction to integer linear programming. Interestingly, the work in Denis and Baldridge (2007) jointly addresses the problems of anaphoricity and co-reference via a *manual* formulation of an integer linear program.

---

7. A situation where a player in CTF moves through the campus at a speed of 100 mph and on her way passes an enemy player is certainly anomalous (and probably caused by GPS sensor noise), but we do not want to say that it is a failed attempt at capturing.

Joint activity modeling has also been shown to yield better recognition accuracy, as compared to "pipeline" baselines as well as baselines that make strong inter-activity independence assumptions. The work of Wu, Lian, and Hsu (2007) performs joint learning and inference over concurrent single-agent activities using a factorial conditional random field model. Similarly, the work of Helaoui, Niepert, and Stuckenschmidt (2010) models interleaved activities in Markov logic. They distinguish between foreground and background activities and infer a time window in which each activity takes place from RFID sensory data. By contrast, we focus on joint reasoning about multi-agent activities and attempts in a fully relational—and arguably significantly more noisy—setting.

The work of Manfredotti, Hamilton, and Zilles (2010) propose a hierarchical activity recognition system formulated as learning and inference in relational dynamic Bayesian networks. Their model jointly leverages observed interactions with individual objects in the domain and the relationships between objects. Since their method outperforms a hidden Markov model by a significant margin, it contributes additional experimental evidence that a relational decomposition of a domain improves model quality.

The work of Landwehr, Gutmann, Thon, Philipose, and De Raedt (2007) casts single-agent activity recognition as a relational transformation learning problem, building on transformation-based tagging from natural language processing. Their system induces a set of transformation rules that are then used to infer activities from sensory data. Since the transformation rules are applied adaptively, at each step, the system leverages not only observed data, but also currently assigned labels (inferred activities). However, the transformation rules are learned in a greedy fashion and experiments show that the model does not perform significantly better than a simple HMM. On the other hand, their representation is quite general, intuitive, and extensible. As we will see, our Markov logic model has a similar level of representational convenience while performing global—instead of greedy—optimization in a significantly more complex domain.

The denoising component of our model can be formulated as a tracking problem. Prior work proposed a relational dynamic Bayesian network model for multi-agent tracking (Manfredotti & Messina, 2009). Their evaluation shows that considering relationships between tracked entities significantly improves model performance, as compared to a nonrelational particle filter baseline. By contrast, our work explores *joint* tracking and activity recognition. However, each GPS reading is annotated with the identity of the corresponding agent. The work of Manfredotti and Messina suggests that our model can be generalized, such that the associations between GPS and agent identities are *inferred* and need not be observed.

Our Markov logic theory can be viewed as a template for a conditional random field (Lafferty, 2001), an undirected graphical model that captures the *conditional* probability of hidden labels given observations, rather than the *joint* probability of both labels and observations, as one would typically do in a directed graphical model. In the relational world, directed formalisms include relational Bayesian networks (Jaeger, 1997) and their dynamic counterparts (Manfredotti, 2009), probabilistic relational models (Koller, 1999; Friedman, Getoor, Koller, & Pfeffer, 1999), Bayesian logic programs (Kersting & De Raedt, 2000), and first-order conditional influence language (Natarajan, Tadepalli, Altendorf, Dietterich, Fern, & Restificar, 2005). Conditional random fields have been extensively applied to activity recognition, and their superior labeling performance over generative models has been demonstrated in a number of both single-agent and multi-agent domains (Liao et al., 2005; Limketkai, Fox, & Liao, 2007; Vail, 2008; Vail & Veloso, 2008; Hu et al., 2008). Since MLNs are often solved as propositionalized CRFs, and the directed alternatives can be compiled into a Bayesian network, it can be expected that discriminative relational models generally outperform their generative counterparts on labeling tasks. However, more work needs to be done to answer this question in its entirety.

Since Markov logic is based on, and in fact subsumes, finite first-order logic, we immediately gain access to a number of techniques developed in the rich field of traditional logic. Current Markov logic solvers take advantage of the underlying logical structure to perform more powerful optimizations, such as Alchemy's lifted inference in belief propagation and MC-SAT (Poon & Domingos, 2006). Additionally, domain pruning, where one uses hard constraints to infer reduced domains for predicates, has been shown to lead to significant speed-ups (Papai, Singla, & Kautz, 2011).

We also leverage this relationship between Markov and first-order logic when inducing an augmented model. Furthermore, presence of dependency cycles introduces additional problems in directed graphical (relational) models. Thus, the fact that, in Markov logic, knowledge can be expressed as weighted first-order formulas combined with the above factors make it a powerful framework best suited for the multi-agent reasoning tasks considered in this work.

Traditional hidden Markov models operate over an alphabet of unstructured (i.e., "flat") symbols. This makes relational reasoning difficult, as one has to either propositionalize the domain, thereby incurring combinatorial increase in the number of symbols and model parameters, or ignore the relational structure and sacrifice information. Logical hidden Markov models (LHMMs) have been proposed to address this problem (Kersting, De Raedt, & Raiko, 2006). LHMMs are a generalization of standard HMMs that compactly represents probability distributions over sequences of logical atoms rather than flat symbols. LHMMs have been proven strictly more powerful than their propositional counterparts (HMMs). By applying techniques from logic-based reasoning, such as unification, while leveraging the logical structure component of the model, Kersting et al. show that LHMMs often require fewer parameters and achieve higher accuracy than HMMs.

LHMMs have been recently applied to activity recognition. In the context of intelligent user interfaces, the work of Shen (2009) designs and evaluates a LHMM model for recognition of people's activities and workflows carried out on a desktop computer. Other researchers proposed a hierarchical extension of LHMMs along with an efficient particle filter-based inference method, and apply it to activity recognition problems in synthetic domains (Natarajan, Bui, Tadepalli, Kersting, & Wong, 2008). Both lines of work show that LHMMs can be learned and applied efficiently, and perform better than plain HMMs.

However, LHMMs are a generative model and therefore are not ideal for pure labeling and recognition tasks, where we typically do not want to make strong independence assumptions about the observations, nor do we want to explicitly model dependencies in the input space. TildeCRF—a relational extension of traditional conditional random fields—has been introduced to address this issue (Gutmann & Kersting, 2006). TildeCRF allows discriminative learning and inference in CRFs that encode sequences of logical atoms, as opposed to sequences of unstructured symbols. TildeCRF specifically focuses on efficient learning of models of sequential data via boosting, and is subsumed by Markov logic, which can produce both discriminative and generative models. We cast our model in the latter framework to make it more general, extensible, and interpretable.

PRISM, a probabilistic extension of Prolog, has been shown to subsume a wide variety of generative models, including Bayesian networks, probabilistic context-free grammars, HMMs (along with their logical extension) (Sato & Kameya, 2001, 2008). However, since the focus of PRISM is on representational elegance and generality, rather than scalability, the sheer size of the state space and complexity of our CTF domain precludes its application here.

Finally, our Markov logic theory augmentation process is related to structure learning, transfer learning, and inductive logic programming. In fact, Algorithm 1 implements a special case of structure learning, where we search for a target theory that explains the training data well, while our declarative bias forces the target theory to differ from the source theory only as much as necessary. Again, with the intuition that failed attempts are similar to their failed counterparts. A number of researchers have focused on structure learning specifically in Markov logic networks. This includes early work on top-down structure learning, where clauses in the knowledge base are greedily modified by adding, flipping, and deleting logical literals (Kok & Domingos, 2005). This search is guided by the likelihood of the training data under the current model. The work of Mihalkova and Mooney (2007) exploit patterns in the ground Markov logic networks to introduce a bottom-up declarative bias that makes their algorithm less susceptible to finding only local optima, as compared to alternative greedy methods. Similarly, the work of Kok and Domingos (2009) introduce a bottom-up declarative bias based on lifted hypergraph representation of the relational database. This bias then guides search for clauses that fit the data. Since the hypergraph is lifted, relational path finding tractable. Interesting work on predicate invention applies relational clustering technique formulated in second-order Markov logic to discover new predicates from relational databases (Kok & Domingos, 2007). The above systems are capable of modeling relatively rich family of logical formulas. Other approaches perform discriminative structure learning and achieve excellent results, but focus on a restricted set of types of formulas (e.g., Horn clauses) (Huynh & Mooney, 2008; Biba, Ferilli, & Esposito, 2008). The work of Davis and Domingos (2009) successfully uses second-order Markov logic in deep transfer learning. They lift the model of the source domain to second-order ML and identify high-level structural patterns. These subsequently serve as declarative bias for structure learning in the target domain.

By its very nature, the inductive logic programming discipline has extensively studied structure learning in deterministic, as well as probabilistic settings (e.g., Muggleton, 2002; De Raedt, 2008; De Raedt, Frasconi, Kersting, & Muggleton, 2008). In fact, our theory augmentation algorithm can be viewed as an efficient Markov logic based version of theory refinement, a well-established ILP technique that aims to improve the quality of a theory in terms of simplicity, fit to newly acquired data, efficiency or other factors (Wrobel, 1996).

Our approach differs from all this work in three main points. First, our declarative bias is defined implicitly by the seed theory of successful activities. Therefore, our theory augmentation algorithm is not limited to any hard-wired set of formula types it can consider. Rather, the search space is defined at run time by extracting motifs from the seed theory. The second distinction lies in computational tractability and exactness of the results. By distinguishing between soft and hard formulas, we are able to search through candidate formulas in a systematic, rather than greedy manner. Consequently, our final learned model requires fewer parameters, which is especially important when the amount of training data is relatively small. Additionally, our weight learning does not experience cold starts, as we leverage the seed theory. The final difference is that, to our knowledge, we are the first to explore structure learning in the context of interplay of success and failure, and their relationship to the intended goals of people's actions.

## 8. Conclusions

This paper took on the task of understanding the game of capture the flag from GPS data as an exemplar of the general problem of inferring human interactions and intentions from sensor data. We have presented a novel methodology—cast in Markov logic—for effectively combining data denoising with higher-level relational reasoning about a complex multi-agent domain. Specifically, we have demonstrated that given raw and noisy data, we can automatically and reliably detect and recognize both successful and failed interactions in adversarial as well as cooperative settings. Additionally, we have shown that success, failure, and the goal of an activity are intimately tied together and having a model for successful events allows us to naturally learn models of the other two important aspects of life. Specifically, we have demonstrated that the intentions of rational agents are automatically discovered in the process of resolving inconsistencies between a theory that models successful instances of a set of activities and examples of failed attempts at those activities.

We have formulated four research questions and designed experiments within the CTF domain that empirically answer them. Compared to alternative approaches to solving the multi-agent activity recognition problem, our augmented Markov logic model, which takes into account not only relationships among individual players, but also relationships among activities over the entire length of a game, although computationally more costly, is significantly more accurate on real-world data. Furthermore, we have illustrated that explicitly modeling unsuccessful attempts boosts performance on other important recognition tasks.

## 9. Future Work

Multi-agent activity recognition is especially interesting in the context of current unprecedented growth of on-line social networks—in terms of their size, popularity, and their impact on our "off-line" lives. In this paper, we show that location information alone allows for rich models of people's interactions, but in the case of on-line social networks, we additionally have access to the content of users' posts and both the explicit and the implicit network interactions. For instance, our recent study shows that, interestingly, about 30% of Twitter status updates reveal their authors' location (Sadilek, Kautz, & Bigham, 2012). These data sources are now available to machines in massive volumes and at ever-increasing real-time streaming rate. We note that a substantial fraction of posts on services such as Facebook and Twitter talk about everyday activities of the users (Naaman, Boase, & Lai, 2010), and this information channel has become available to the research community only very recently. Thus, if we are able to reason about human behavior and interactions in an automated way, we can tap the colossal amounts of knowledge that is—at present—distributed across the whole population.

We are currently extending our model to handle not only explicit GPS traces, but also be able to *infer* the location of people who do not broadcast their GPS coordinates. The basic idea is, again, to leverage the structure of relationships among people. The vast majority of us participate in on-line social networks and typically some of our friends there do publish their location. We thus view the GPS-enabled people as noisy location sensors and use the network interactions and dynamics to estimate the location of the rest of the users. At present, we are testing this approach on public tweets.

## Acknowledgments

We thank anonymous reviewers for their constructive feedback. We further thank Sebastian Riedel for his help with theBeast, and to Radka Sadíková and Wendy Beatty for their helpful comments. This work was supported by ARO grant #W911NF-08-1-0242, DARPA SBIR Contract #W31P4Q-08-C-0170, and a gift from Kodak.